\documentstyle[12pt,epsf]{article}
\def\href#1#2{#2}   
\newif\ifdraft
\draftfalse	  
\reversemarginpar   
\makeatletter
\let\mlabel=\label
\let\adkendequation=\endequation%
\def\endequation{\adkendequation\adklabel\global\@ignoretrue}
\let\adkendeqnarray=\endeqnarray%
\def\endeqnarray{\adkendeqnarray\adklabel\global\@ignoretrue}
\newbox\marglabbox
\def\adklabel{\ifvoid\marglabbox\else\marginpar{\unhbox\marglabbox}\fi}
\def\label#1{\ifdraft\ifmmode%
  \global\setbox\marglabbox=\hbox{\hfill\fbox{\tiny\verb*~#1~}}%
  \else\ifinner\else\marginpar{\hfill\fbox{\tiny\verb*~#1~}}%
  \fi\fi\fi \mlabel{#1}}
\makeatother
\ifdraft%
\fi
\font\twelvebb=msbm12
\font\tenbb=msbm10
\font\sevenbb=msbm7
  \newfam\bbfam
  
  \textfont\bbfam=\twelvebb
  \scriptfont\bbfam=\tenbb
  \scriptscriptfont\bbfam=\sevenbb
\font\twelveeusm=eusm10 scaled 1200
\font\teneusm=eusm10
  \newfam\eusmfam
  
  \textfont\eusmfam=\twelveeusm
  \scriptfont\eusmfam=\teneusm
  \scriptscriptfont\eusmfam=\scriptfont\eusmfam
\font\twelvefrak=eufm10 scaled 1200
\font\tenfrak=eufm10
  \newfam\frakfam
  
  \textfont\frakfam=\twelvefrak
  \scriptfont\frakfam=\tenfrak
  \scriptscriptfont\frakfam=\scriptfont\frakfam
\def\sqr#1#2{{\vcenter{\hrule height.#2pt
   \hbox{\vrule width.#2pt height#1pt \kern#1pt
      \vrule width.#2pt}
   \hrule height.#2pt}}}

\def\bsqr#1#2{{\vrule width #1pt height#2pt}}
\def\bsquare{{\mathchoice\bsqr66\bsqr66\bsqr33\bsqr33}}
\def\badbreak{\penalty1000}

\newcommand{\gfive}{\gamma_{5}}             
\newcommand{\cP}{{\cal P}}                  
\newcommand{\cS}{{\cal S}}                  
\newcommand{\cml}{{\cal C}}                 
\newcommand{\lsz}{{\ell}}                   
\newcommand{\lfd}{{\varphi}}                
\newcommand{\lin}{{\mu}}                    

\begin{document}

\begin{center}
{\Large{\bf The Analysis of Space-Time Structure in QCD}} \\
\vspace*{.1in}
{\Large{\bf Vacuum I: Localization vs Global Behavior in}}\\
\vspace*{.1in}
{\Large{\bf Local Observables and Dirac Eigenmodes}}\\
\vspace*{.4in}
{\large{Ivan Horv\'ath 
}}\\
\vspace*{.15in}
University of Kentucky, Lexington, KY 40506\\

\vspace*{0.2in}
{\large{October 27 2004}}

\end{center}

\vspace*{0.15in}

\begin{abstract}
  \noindent 
  The structure of QCD vacuum can be studied from first principles using the lattice-regularized 
  theory. This line of research entered a qualitatively new phase recently, wherein the 
  space-time structure (at least for some quantities) can be directly observed in configurations 
  dominating the QCD path integral, i.e. without any subjective processing of typical 
  configurations. This approach to QCD vacuum structure does not rely on any proposed picture 
  of QCD vacuum but rather attempts to characterize this structure in a model-independent 
  manner, so that a coherent physical picture of the vacuum can emerge when such unbiased
  numerical information accumulates to a sufficient degree. An important part of this program
  is to develop a set of suitable quantitative characteristics describing the space-time 
  structure in a meaningful and physically relevant manner. One of the basic pertinent issues
  here is whether QCD vacuum dynamics can be understood in terms of localized vacuum objects,
  or whether such objects behave as inherently global entities. The first direct studies 
  of vacuum structure strongly support the latter. In this paper, we develop a formal 
  framework which allows to answer this question in a quantitative manner. We discuss in detail 
  how to apply this approach to Dirac eigenmodes and to basic scalar and pseudoscalar composites
  of gauge fields (action density and topological charge density). The approach is illustrated 
  numerically on overlap Dirac zero modes and near-zero modes. This illustrative data provides 
  direct quantitative evidence supporting our earlier arguments for the global nature of QCD 
  Dirac eigenmodes.
\end{abstract}

\section{Introduction}

The studies of QCD vacuum structure using lattice regularization have a long history. However,
only recently did this research took on the challenge to approach the problem independently,
without the guidance from a particular idealized model of the QCD 
vacuum~\cite{Hor02B,Hor03A,Hor01A,deG01A,Hor02A}. In particular, there was a significant progress 
in the quest to avoid any subjective processing of configurations dominating the regularized 
QCD path integral. This progress happened at two qualitatively different levels. First, it was 
recognized that the {\em local} behavior of Dirac eigenmodes can serve as an ideal unbiased 
probe of the underlying vacuum structure~\cite{Hor01A,deG01A,Hor02A}, which lead to an increased
level of activity in this direction~\cite{Followup,Other}. The eigenmode approach, though unbiased, 
is still indirect since certain theoretical framework or model is needed to connect the structure 
of eigenmodes to the structure of underlying fields~\cite{Hor02B}. A purely direct approach, 
wherein one would inspect the structure of local fields directly in regularized QCD ensemble 
is not possible yet in general. However, in a rather surprising turn of events, it was 
demonstrated~\cite{Hor03A} that using topological charge density operator associated with 
the overlap Dirac operator~\cite{overlap}, the fundamental space-time structure in topological 
charge density appears, and the direct approach is thus possible (see also~\cite{Tha04}). 
Moreover, in this framework it is possible to study topological charge fluctuations at arbitrary 
low-energy scale by employing the effective topological field constructed via Dirac eigenmode 
expansion~\cite{Hor02B}. 

The drive for model-independent approach to QCD vacuum structure requires the introduction of 
model-independent concepts in order to characterize this structure. The goal of this paper is 
to propose a sufficiently general framework designed for systematic studies focusing on the
questions of space-time localization and/or global behavior in QCD vacuum. Since the seminal
work of Anderson~\cite{And58}, the issues of localization played an important role in solid 
state physics. Here one usually means the exponential localization in the wave function 
describing an electron in the solid. Localized wave-function is pinned to a point $x_0$ and 
behaves so that $|\psi(x)| \le c e^{-(x-x_0)/\xi_0}$, where the maximal $\xi_0$ is 
the localization scale. In the localized regime the pinning points are dilute in 3-d space 
and the wave-function effectively has a support on non-overlapping localized regions. When 
increasing the amount of disorder the localized regions start to overlap, and the modes can 
become delocalized if the linear size of (at least some) path-connected subsets of the support 
becomes comparable to the size of the sample. The physics of the system changes drastically 
when that happens. In the case of QCD vacuum structure the situation is unfortunately more 
complicated for several reasons. (1) We cannot assume the existence of exponentially localized 
structures neither in local composite fields nor in Dirac eigenmodes. Indeed, there is no direct 
evidence that exponential localization takes place in QCD vacuum. 
(2) In case of certain proposed topological structures 
the standard nomenclature already violates the above classification. For example, a single 
instanton in arbitrary large volume and the associated 't Hooft mode would both be classified 
as global, and yet they are referred to as a localized structure which can become ``delocalized'' 
for interacting multi-instanton case. In fact, the nomenclature (especially in lattice 
QCD vacuum literature) is quite loose and it is a common practice to refer to almost any peak 
(or just a lower-dimensional section that resembles a peak) freely as localized structure. 
(3) There is no obvious tunable parameter in QCD (analogous to the amount of disorder in 3-d 
solids) which could be used to identify the potential candidates for localized objects, which 
in turn could be used for at least approximate physical analysis of possible delocalization.
\footnote{Instantons could in principle play the role of such building blocks, but there is 
a considerable numerical evidence that they can not be identified neither in Dirac 
eigenmodes~\cite{Hor02A} nor in effective densities~\cite{Hor02B}. Instantons are thus unlikely 
to provide a suitable starting point (``basis'') for the description of global behavior.} 
(4) It has been suggested that vacuum structure in topological density~\cite{Hor03A} 
(and the structure reflected in Dirac near-zero modes~\cite{Hor02A}) is not localized 
and cannot even be viewed as delocalized. Rather this structure is {\em inherently global 
(super-long-distance)} in the sense that one can not identify any local parts contained in 
bounded regions of space-time that would be physically dominant over the rest of 
the global structure. 

Given the above, the challenge we face in this work is how to meaningfully decide the question
whether QCD vacuum structure is localized, delocalized or inherently global without making any
a priori assumptions about the nature of the structure. To answer this question unambiguously is 
clearly of great importance for our views of QCD vacuum and for the possible directions of future 
research. For example, if the topological structure is inherently global as argued in~\cite{Hor03A}, 
then the attempts to understand topological charge fluctuations in terms of lumps concentrated in 
bounded regions of space-time will not be productive. One will have to approach the problem
in terms of some fundamental global structure instead.

The general approach described here is the result of two considerations. (A) The concept of 
localization relates to the geometric properties of the ``support''. Here support is the
subset of space-time which, from the physics point of view, dominates the behavior of some 
local observable. Assuming that the support can be identified unambiguously, the question 
becomes of purely geometric nature. Indeed, consider the three particular cases schematically 
depicted in Fig.~1. In case (a) the support $\Omega_1 \subset \Omega$ consists of disconnected 
pieces with the characteristic linear size much smaller than the linear size of space-time 
$\Omega$. We associate this with the case of {\em localized} physical structure. In part (b) 
the support exhibits connected pieces whose linear size is comparable to the size of 
the space-time. Such connected pieces are geometrically global and could mediate physics 
over very large distances. However, the localized parts can still be recognized within 
the geometrically global component, and there might also be isolated pieces that still retain 
their localized identity. We associate this with {\em delocalized} physical structure. Finally, 
in part (c) the behavior is dominated by geometrically global pieces without any distinguishable 
localized parts that could be used for valid physical analysis. The structure is 
{\em inherently global} and can mediate physics over arbitrarily large distances.
\footnote{For this reason we sometimes refer to such structure as 
          {\em super-long-distance}~\cite{Hor03A}.}
In Sec.~\ref{WDS} we will define geometric characteristics that will help to identify these 
typical situations.
(B) Without the existence of a strict support
\footnote{The function $f$ on domain $\Omega$ is said to have a strict support on 
          $\Omega_1 \subset \Omega$ if $f(x)\ne 0$ for $x\in \Omega_1$ and $f(x)=0$ for 
          $x\in\Omega-\Omega_1$.} 
and without exponential localization, the very notion of support is ambiguous. Indeed, it is 
possible to give many generic definitions with very different results for what the support is. 
Therefore, rather than fixing the definition, we advocate the use of 
{\em space-time cumulative function} for the local quantity studied. Ranking the space-time 
points by the magnitude of local quantity in question, this function embodies the notion of 
support and contains detailed information on how much different parts of space-time contribute 
to the associated physics. In this way we can study the issues of localization as described 
in (A) for different levels of saturation. In Sec.~\ref{cumul} we describe this idea in detail 
and set up a general framework that will allow us to decide unambiguously the issue of 
localization in QCD vacuum.
 
In Sec.~\ref{cases} we discuss the application of the above ideas specifically to Dirac 
eigenmodes (both zero modes and near-zero modes), and the basic scalar and pseudoscalar 
composites of the gauge fields (action density and topological charge density). We illustrate 
our approach explicitly (including some numerical results) on zero modes and near-zero modes 
of the overlap operator~\cite{overlap}. These examples show, within the formal framework 
developed here, that low-lying Dirac eigenmodes are inherently global in QCD as first 
observed in~\cite{Hor02A}. The full description of these results will be given in the upcoming 
publication~\cite{gl_modes}.

   \begin{figure}
   \begin{center}
      \centerline{
      \epsfxsize=0.33\hsize\epsffile{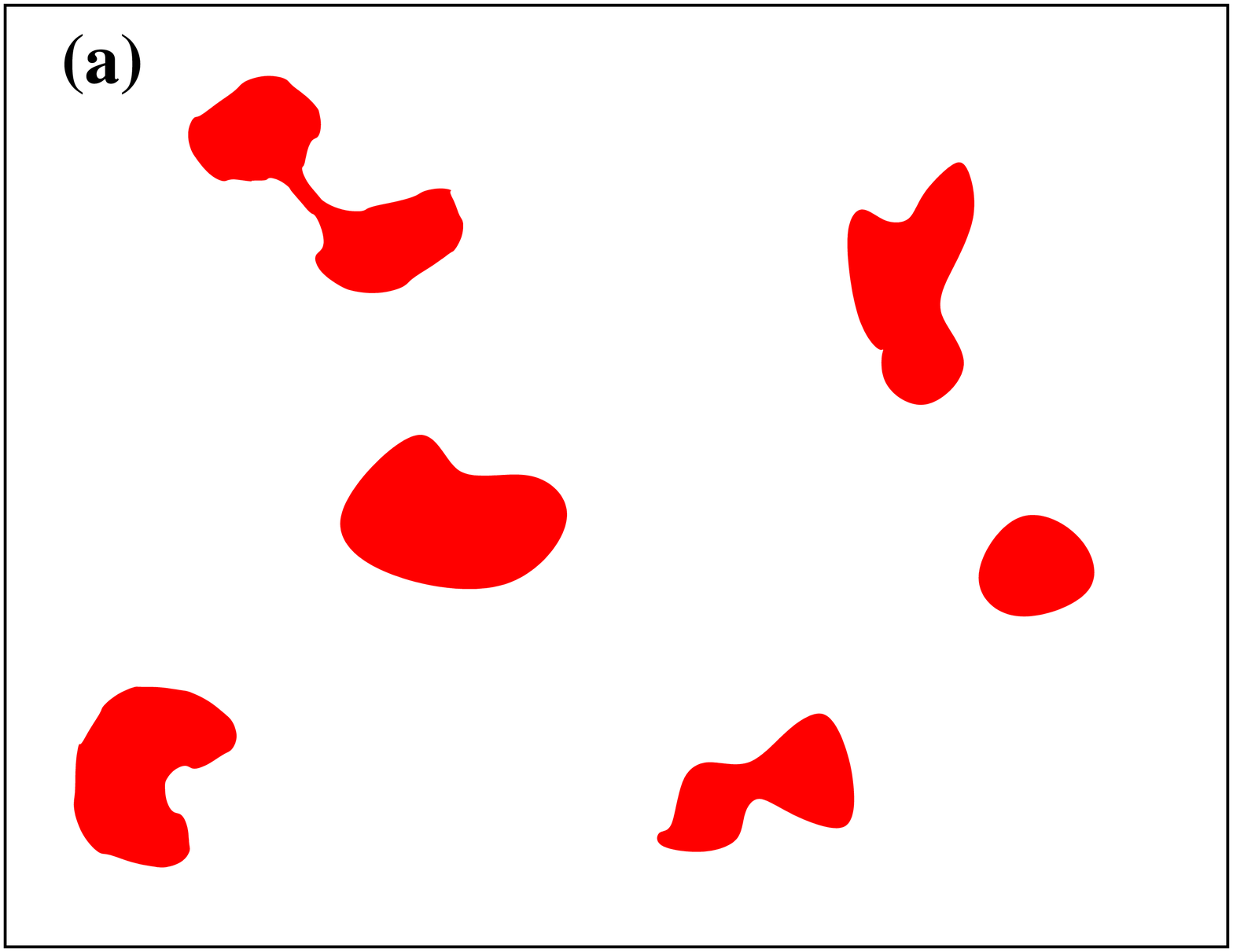}
      \epsfxsize=0.33\hsize\epsffile{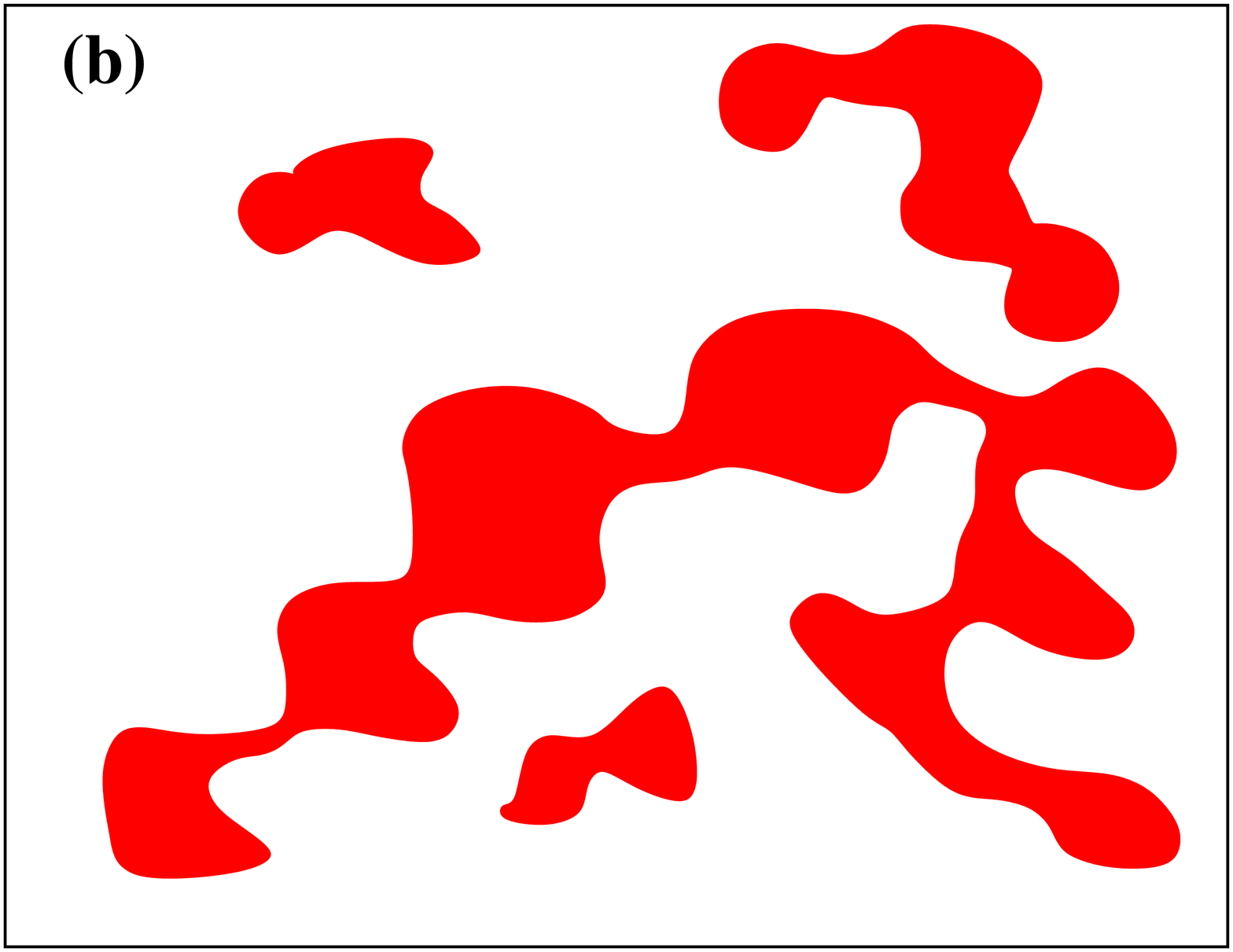}
      \epsfxsize=0.33\hsize\epsffile{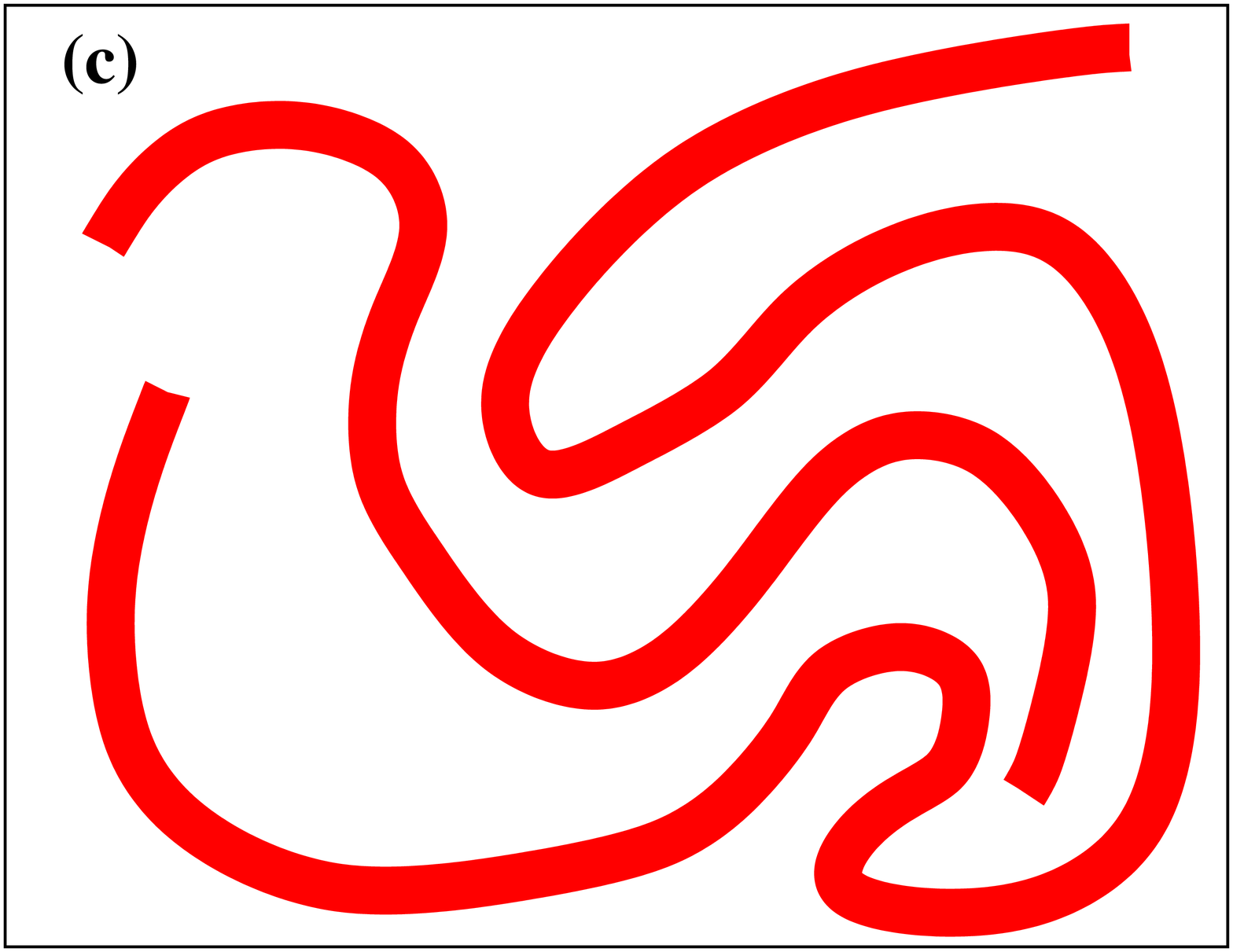}
      }
      \caption{The examples of localized (a), delocalized (b), and inherently global (c)
      structure.}
   \label{cartoon1}
   \end{center}
   \end{figure}

\section{The case of well-defined support}
\label{WDS}

In this section we will assume that for every typical gauge background we were given a subset 
$\cS^\lfd$ of space-time $\Omega$ which dominates the behavior of some local physical quantity 
$\lfd(x)$ defined on $\Omega$. Such support $\cS^\lfd$ could have been determined via exponential 
localization or by using some physical input which helped to select it uniquely for typical 
configurations contributing to regularized QCD path integral. It should be noted that when we 
talk about ``space-time'' we implicitly have in mind a finite hypercubic discretization of 
the torus with inherited Euclidean metric. However, the considerations of this section are 
independent of this assumption, and one can equally well think of the Euclidean torus itself. 
To proceed, it will be useful to first define the notion of space-time geometric structure 
as it will be used here and in some other studies of the subject. 

\subsection{Geometric structure and its global behavior}
\label{gstruc}

We adopt a very general definition of geometric structure which can be easily refined when 
used in more specific situations. It should be emphasized that the discussion in this subsection
is purely geometrical and independent from the physics context in which it will be applied.

\medskip
\noindent{\em \underline{Definition\,}($\,$Geometric Structure$\,$):$\;$}
     {\em Let $\Omega$ be the set of space-time points. The collection  
     ${\rm S} = \{\cS_i,\, i=1,\ldots,N\,\}$ of path-connected subsets $\cS_i \subset \Omega$ 
     such that $\cS_i \cup \cS_j$ is disconnected for all $i \ne j$ will be referred to as 
     a geometric space-time structure.}
\smallskip

\noindent In case of hypercubic lattice the term path-connectivity means the usual link 
connectivity. The reason why the above definition focuses on path-connected sets is the
underlying expectation that in physically relevant situations such sets will be associated 
with mediation of long-distance physics.
\smallskip

We wish to characterize the degree of global behavior in geometric structure S via that of 
its connected components $\cS_i$. To do that, we need a notion of ``linear size'' for  
connected components regardless of their shapes. This is standard for arbitrary subsets of metric 
spaces and is defined as a maximum (supremum) over distances between any two points of the set.

\medskip
\noindent{\em \underline{Definition\,}($\,$Linear Size$\,$):$\;$}
     {\em Let $\Gamma$ be the subset of the Euclidean space-time $\Omega$. The linear 
     size of $\Gamma$ is defined as $\lsz(\Gamma) \equiv \max \{\, |x-y| : \; x,y \in \Gamma \,\}$.} 
\medskip

\noindent The subset $\Gamma$ of space-time will be called {\em global} if its linear size is 
          comparable to the size of the underlying space, i.e. $\lsz (\Gamma) \approx \lsz(\Omega)$. 
          Conversely, $\Gamma$ will be called {\em localized} if $\lsz(\Gamma) \ll \lsz(\Omega)$.  
\smallskip

To quantify the degree of global behavior in geometric structure S, it is useful to define 
the following two characteristics. 
\footnote{The characteristic of this type has already been introduced in Ref.~\cite{Hor03A}, 
          where its use lead to the proposal that the newly discovered low-dimensional structure 
          in topological charge density is super-long-distance.} 
The first is the maximal linear size $\lsz_{max}({\rm S})$ of the connected component, i.e. 

\begin{equation}
     \lsz_{max}({\rm S}) \;\equiv\; \max_{\cS_i \in {\rm S}}\, \{\, \lsz(\cS_i)\,\}
     \label{eq:10}
\end{equation}

\noindent The second is the point-wise average of the linear size over the points of the structure. 
More precisely, to every point $x\in \cS \equiv \cup_i \cS_i$ we assign the linear size of the
corresponding connected component, 
\footnote{Note that we distinguish between the structure S representing the collection of 
          path-connected sets and the set of all points belonging to the structure 
          $\cS \equiv \cup_i \cS_i$.} 
and take the average over $\cS$

\begin{equation}
     \lsz_{\rm S}(x) \;\equiv\; \lsz(\cS_i), \; x \in \cS_i   \qquad\qquad
     \lsz_{pta}({\rm S}) \,\equiv\, \langle\, \lsz_{\rm S}(x) \,\rangle_{\cS}
     \label{eq:20}
\end{equation}

\noindent Note that $\lsz_{\rm S}(x)$ is a piece-wise constant function on $\cS$. For continuum 
space-time manifold $\Omega$ a precise definition of the integrals entering the mean value has 
to be given for different classes of subsets $\cS_i$. However, for the regularized case of 
discretized torus the above expression has the usual meaning of the point-wise average.
\smallskip

The purpose of introducing $\lsz_{max}({\rm S})$ and $\lsz_{pta}({\rm S})$ is that their 
values can signal different regimes of global behavior in the structure. Indeed, while 
$\lsz_{max}({\rm S})$ indicates whether there is any global behavior within the structure 
at all, $\lsz_{pta}({\rm S})$ measures to what extent is such behavior prevalent in 
the structure as a whole. Obviously,  $\lsz_{pta}({\rm S}) \le \lsz_{max}({\rm S})$ for 
arbitrary S. Let us distinguish three typical situations:
\smallskip

\noindent (a) $\lsz_{pta}({\rm S}) \le  \lsz_{max}({\rm S}) \ll \lsz(\Omega) \;$ 
 {\em (localized structure).} $\;$ In this case all connected components behave as 
 localized subsets. Fig.~1a represents an example of such geometric behavior.
\smallskip 

\noindent (b) $\lsz_{pta}({\rm S}) <  \lsz_{max}({\rm S}) \approx \lsz(\Omega) \;$ 
 {\em (partially global structure).} Here only parts of the structure behave as strictly 
 global, while parts of it can still have localized character. Fig.~1b illustrates this
 case.  
\smallskip

\noindent (c) $\lsz_{pta}({\rm S}) \approx  \lsz_{max}({\rm S}) \approx \lsz(\Omega) \;$ 
 {\em (global structure).} In this case the structure has to be viewed as global since 
 it is completely dominated by global connected parts, as exemplified in Fig.~1c.
\smallskip

\noindent 

In essence, we are proposing to associate with any geometric structure two non-negative 
numbers less than one, namely $\lsz_{pta}({\rm S})/\lsz(\Omega)$ and 
$\lsz_{max}({\rm S})/\lsz(\Omega)$. These values quantify the global behavior 
of the structure. We emphasize that the above classification is purely geometrical and 
will be useful only to the extent to which the characteristic situations described above 
turn out to be typical for space-time structures occurring in the QCD vacuum.

\subsection{Support from physics considerations}
\label{PWDS}

We now come back to the discussion of the situation where the space-time support 
$\cS^\lfd \subset \Omega$ corresponding to local observable $\lfd(x)$ is assumed to be 
uniquely determined on physical grounds. It is easy to see that $\cS^\lfd$ defines 
a space-time structure S$^\lfd$ as defined in the previous subsection. Indeed, arbitrary 
$\cS^\lfd$ can be uniquely decomposed into maximal connected subsets, 
i.e. $\cS^\lfd = \cup_i \cS^\lfd_i$ such that $\cS^\lfd_i$ is connected, and that 
$\cS^\lfd_i \cup \cS^\lfd_j$ is disconnected  for all $i \ne j$. 
Then S$^\lfd = \{\cS_i^\lfd,\, i=1,\ldots,N\,\}$. We can thus use the geometric
concepts of the previous subsection to characterize the degree of its global behavior.
However, since the geometry of S$^\lfd$ has been determined by the underlying physics, 
the global or localized geometric nature of S$^\lfd$ are expected to be manifestations 
of qualitatively very different types of underlying dynamics which we try to understand. 
We are implicitly making the following associations:

\smallskip

\noindent (a) If S$^\lfd$ behaves as localized geometric structure, then we expect 
the vacuum objects driving the dynamics of the gauge field also to be localized, 
with typical size approximately given by $\lsz_{pta}({\rm S}^\lfd)$. From the physics 
point of view we are thus dealing with localized structure. 
\footnote{It is worth emphasizing here that there is a relatively widely-spread 
          confusion in lattice QCD literature about the use of the term {\em localized}.
          In particular, the notion of localization is frequently interchanged with
          the potency of the support. In reality, these two concepts are entirely independent.
          The support can be a set of measure zero and yet the structure can still be 
          global. Some global characteristics (such as $\lsz_{pta}$ and $\lsz_{max}$)
          have to be computed in order to determine localized versus global nature 
          of the structure.} 
\smallskip

\noindent (b) If S$^\lfd$ behaves as partially global geometric structure, it is an
indication that we are dealing with delocalized physical structure. Indeed, the 
presence of global connected subsets signals that the dynamics produces correlations over 
very large distances but, at the same time, the presence of localized subsets points 
to the existence of underlying localized objects that mediate this correlation. 
Such objects can still serve as a useful tool for theoretical analysis of 
the situation.
\smallskip

\noindent (c) If S$^\lfd$ behaves as global geometric structure, then this indicates 
that the underlying dynamics produces correlations that universally span the whole 
system. This could happen as a result of two qualitatively different physical 
situations. The first is the case of extreme delocalization, wherein the underlying 
localized objects are essentially never seen individually, but can still be 
distinguished within the global subsets and are relevant for physics. The second 
possibility is that the global subsets do not contain any privileged parts and behave 
as a single whole. Here the notion of a localized object is no longer useful to analyze 
the physics, and the structure is inherently global (super-long-distance).

\smallskip

To avoid the confusion in nomenclature in what follows, let us summarize. Our goal 
is to be able to distinguish three {\em physically} distinct situations: localized, 
delocalized and inherently global (super-long-distance) behavior. At the same time, 
we introduced the concepts that distinguish three {\em geometric} space-time 
arrangements: geometrically localized, partially global and global structure. If there 
is a well-defined support $\cS^\lfd$ dominating the physical behavior of $\lfd$, 
we associate the localized geometric nature of corresponding S$^\lfd$ with the underlying 
localized physical behavior. Similarly, the partially global nature of S$^\lfd$ is 
associated with delocalization. However, if $\cS^\lfd$ is a global geometric structure, 
we could be dealing with two qualitatively different physical cases, namely extreme 
delocalization or super-long-distance behavior. To distinguish the two, one could 
introduce additional geometric concepts. However, we will take a different approach 
in what follows.

\section{The use of fraction supports}
\label{cumul}

While the discussion of the previous section provides a partial basis for analyzing 
the issues of localization, it is not quite satisfactory. Indeed, the main problem 
is that frequently it is not possible to use it because there is no unique way 
to determine the support for given local quantity $\lfd(x)$ from physics considerations 
or otherwise. Even though there are many ways to assign a support to a function in 
a generic way (i.e. regardless of the functional form), the arbitrariness of such 
procedures makes it dangerous for making definite conclusions. We discuss the related 
issues in the Appendix~\ref{app:1}. 
\smallskip

In order to overcome this shortcoming, we propose to work with many ``potential'' 
supports and monitor both the degree of global behavior, and the degree to which 
$\lfd(x)$ is concentrated in such possible supports. In this way, much more complete 
information enters the consideration. We will argue and demonstrate (both here and 
in upcoming publications) that this strategy leads to definite conclusions in the specific 
case of QCD vacuum structure. The procedure can be described in few steps and, to exemplify 
the discussion, we will frequently refer to the case of Schr\"odinger wave function, 
i.e. $\lfd(x) \longrightarrow \psi(x)$, where $x$ is a space coordinate. In what follows, 
we will assume (both in definitions and formulas) that $\Omega$ is a hypercubic 
discretization of the torus. 
\smallskip

1. Choose a positive semi-definite measure $\lin^\lfd(x)$ that characterizes the physically 
   relevant ``strength'' (intensity) of $\lfd(x)$. For Schr\"odinger wave function this 
   is obviously $\lin^\psi(x)=\psi^{\dagger}(x)\psi (x)$ because this represents 
   the probability density for the particle to be in the vicinity of $x$. The role of 
   $\lin^\lfd(x)$ is to introduce an importance ranking for all space-time points: 
   high-intensity points are physically more relevant than low-intensity points. This 
   allows us to define the family of nested subsets $\cS^\lfd(f)$ (fraction supports) 
   respecting this ranking. 
   \smallskip

   \noindent{\em \underline{Definition\,}(Fraction Support):$\;$ Let $\lfd(x)$ be the local 
   observable on space-time $\Omega$ with intensity $\lin^\lfd(x)$. The collection 
   $\cS^\lfd(f) \subset \Omega$ of highest intensity-ranked space-time points 
   that occupy fraction $f$ of available volume, i.e. $N(\cS^\lfd(f))/N(\Omega) = f$,
   will be referred to as a fraction support corresponding to $f$ and $\lfd(x)$.}
   \smallskip

   \noindent In the above definition the ``volume'' $N(\Gamma)$ for arbitrary 
   $\Gamma \subset \Omega$ represents the number of lattice points contained in $\Gamma$.
   The sets $\cS^\lfd(f)$ form a family of potential supports whose global behavior will 
   be examined. 
   \smallskip

2. While the volume fraction $f$ is a convenient label for fraction supports, we also
   need a measure expressing how good a support $\cS^\lfd(f)$ would be, i.e. to which
   degree the behavior of $\lfd(x)$ within $\cS^\lfd(f)$ dominates relative to 
   the whole $\Omega$. In case of Schr\"odinger particle the appropriate measure is 
   the total probability of the particle to be within the support, i.e. the sum
   of $\lin^\psi(x)=\psi^\dagger(x)\psi(x)$ over $\cS^\psi(f)$. In general, this 
   information is contained in the normalized cumulative function of intensity
   \footnote{To arbitrary $\Gamma \subset \Omega$ and $\lfd(x)$ we can assign a 
   coefficient $\sum_\Gamma \, \lin^\lfd(x)/\sum_{\Omega}\, \lin^\lfd(x)$ expressing
   its quality as a potential support. $\cml^\lfd(f)$ represents a maximum of this 
   coefficient over all subsets occupying the volume fraction $f$ of the underlying 
   space.}

   \begin{equation}
      \cml^\lfd(f) \,\equiv \, {\;\sum\limits_{\cS^\lfd(f)}^{}\, \lin^\lfd(x)\;}/
                               {\;\sum\limits_{\Omega}^{}\, \lin^\lfd(x)\;}               
      \label{eq:30}
   \end{equation}

   \noindent At the regularized level considered here (before taking the continuum limit) $f$ 
   is a discrete variable chosen from the set 
   $\{\, n/N(\Omega),\, n=0,1,2,\ldots,N(\Omega)\,\}\subset [0,1]$. 
   However, we will keep the ``continuous'' notation and associate $\cml^\lfd(f)$ with 
   a continuous piece-wise linear function connecting the points of the discrete map. 
   In the continuum limit (and as a regularized approximation), $\cml^\lfd(f)$ is 
   a non-decreasing concave function from interval $[0,1]$ to itself, such that $\cml^\lfd(1)=1$. 
   If $\lin^\lfd(x)$ is non-zero only in a single point or on the set of measure zero, 
   then $\cml^\lfd(f)=1$ identically. In the other extreme situation when $\lin^\lfd(x)$ is
   constant through space-time, we have $\cml^\lfd(f)=f$. In case of arbitrary 
   strict support, i.e. when there exists $\Gamma \subset \Omega$ such that 
   $\lin^\lfd(x)\ne 0$ for $x\in \Gamma$ and $\lin^\lfd(x)= 0$ for $x\not\in \Gamma$,
   the function $\cml^\lfd(f)$ reaches value $1$ at the corresponding fraction $f_\Gamma$ 
   and remains constant on the interval $[f_\Gamma,1]$. 

   \medskip  

   \noindent We emphasize that $\cml^\lfd(f)$ is a basic characteristic of the 
   space-time structure induced by $\lfd(x)$. It has a specific behavior for particular 
   classes of space-time distributions and contains detailed quantitative information 
   about how much is $\lfd(x)$ concentrated in fraction supports $\cS^\lfd(f)$. 
   We propose that this information is necessary in order to judge the potential global 
   character of the physically relevant structure.
   \smallskip

   3. Every fraction support $\cS^\lfd(f)$ generates a corresponding geometric 
   structure S$^\lfd(f)$ via decomposition into maximal connected subsets. To monitor
   the global behavior of these structures, we compute the corresponding characteristics 
   introduced in Sec.~\ref{gstruc}, and thus define the functions 
   \begin{equation}
      \lsz_{max}^\lfd(f) \equiv \lsz_{max}({\rm S}^\lfd(f))  \qquad\qquad
      \lsz_{pta}^\lfd(f) \equiv \lsz_{pta}({\rm S}^\lfd(f))
      \label{eq:40}
   \end{equation}

   When measuring the length in the units of maximal linear distance $\lsz(\Omega)$
   (which we will assume from now on), then  $\lsz_{max}^\lfd(f)$ and $ \lsz_{pta}^\lfd(f)$
   are functions from interval $[0,1]$ to itself and 
   $\lsz_{max}^\lfd(1) = \lsz_{pta}^\lfd(1) = 1$ similarly to $\cml^\lfd(f)$
   (see however Sec.~\ref{agrad}). Contrary to $\cml^\lfd(f)$ though, these functions
   are not necessarily concave. Also, $\lsz_{max}(f)$ is a non-decreasing function,
   while the monotonic properties of $\lsz_{pta}(f)$ are indefinite in general.
   \smallskip

   \noindent We have thus defined three characteristic functions associated with local 
   physical observable $\lfd(x)$ defined on space-time $\Omega$. While $\cml^\lfd(f)$ 
   describes and quantifies the tendency of $\lfd(x)$ to be concentrated in preferred
   regions of space-time, $\lsz_{max}^\lfd(f)$ and $\lsz_{pta}^\lfd(f)$ describe and 
   quantify the degree of global behavior for such preferred regions. The knowledge
   of these functions allows us to judge the degree of global behavior for arbitrary
   degree of saturation in physically relevant intensity $\lin^\lfd(x)$. We propose 
   that this detailed information has to be considered in conjunction in order to make 
   definite conclusions on whether the physical structure inducing the behavior of 
   $\lfd(x)$ is localized, delocalized, inherently global or indefinite. Following 
   the discussion in Sec.~\ref{WDS}, we distinguish three typical and qualitatively 
   different types of behavior shown schematically in Fig.~2.  
   \smallskip

  \begin{description}
   \item[(i)] $\cml^\lfd(f)$ saturates considerably faster than $\lsz_{max}^\lfd(f)$ 
     as exemplified in Fig.~2a. 
     \footnote{Note that it is not essential that we have drawn the concave graphs 
               for $\lsz_{max}^\lfd(f)$ and $\lsz_{pta}^\lfd(f)$ in Fig~2a. These 
               functions could in principle be convex or mixed as well.}
     This means that at fractions where $\cml^\lfd(f)$ starts to saturate (smallest 
     fractions such that   
     $\cml^\lfd(f)\:\raisebox{-0.8ex}{$\stackrel{\raisebox{-1.5ex}{$\textstyle <$}}
     {\sim}$}\:1$) 
     the corresponding fraction supports form a localized geometric structure, i.e. 
     $\lsz_{max}^\lfd(f) \ll \cml^\lfd(f)$. In this case we classify the underlying 
     physical structure as localized since the dominating fields are concentrated 
     on such geometric structure.

   \item[(ii)] $\cml^\lfd(f)$ saturates at similar fractions as either 
     $\lsz_{max}^\lfd(f)$ or $\lsz_{pta}^\lfd(f)$, or in between the two, as 
     exemplified in Fig.~2b. This situation indicates that the underlying physical 
     structure is delocalized. Indeed, if $\cml^\lfd(f)$ saturates faster than 
     $\lsz_{pta}^\lfd(f)$ then the fraction support at the point of saturation 
     forms a partially global geometric set, i.e. $\lsz_{max}^\lfd(f) \approx
     \cml^\lfd(f) > \lsz_{pta}^\lfd(f)$, which is typical of delocalization 
     (see Sec.~\ref{PWDS}). On the other hand, if $\cml^\lfd(f)$ and 
     $\lsz_{pta}^\lfd(f)$ saturate at approximately the same rate, then we are
     dealing with geometrically global structure at the point of saturation, which 
     is common both to extremely delocalized and inherently global cases. However, 
     the signature of extremely delocalized structure is that a small decrease in fraction 
     changes global geometric behavior to partially global, where some individual localized 
     objects starts to appear in isolation. This is precisely what happens in this case.  
     
   \item[(iii)] $\cml^\lfd(f)$ saturates considerably slower than $\lsz_{pta}^\lfd(f)$, 
     as shown in Fig.~2c. This means that at fractions where $\cml^\lfd(f)$ starts to 
     saturate the corresponding fraction supports form a geometrically global structure, 
     i.e. $\lsz_{pta}^\lfd(f) \approx \cml^\lfd(f)$. Moreover, this arrangement is very 
     stable with respect to the change of the fraction. This indicates that the underlying 
     physical structure is inherently global (super-long-distance). 
 
 \end{description}

  We emphasize that a generic space-time arrangement does not necessarily have to fall
  into any of the typical cases described above. If that happens, than we have an 
  indefinite situation which has to be analyzed in more detail by other means.

    \begin{figure}
    \begin{center}
      \centerline{
      \epsfxsize=16.5truecm\epsffile{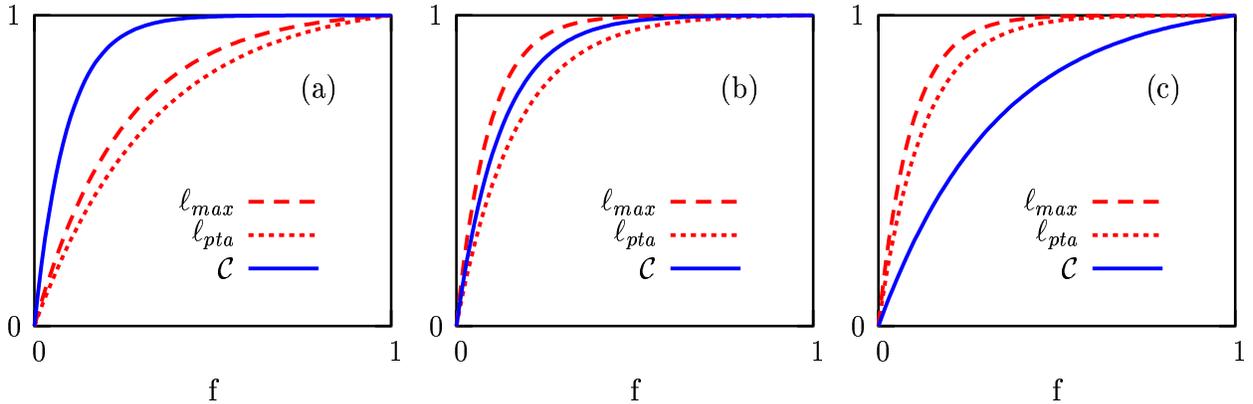}
      }
      \caption{The signatures of localized (a), delocalized (b), and inherently global (c)
               behavior.}
      \label{cartoon2}
      \vskip -0.3in
    \end{center}
    \end{figure}

   \smallskip
   4. While the fraction $f$ of space-time is a convenient variable to introduce 
   the functions $\cml^\lfd(f)$, $\lsz_{max}^\lfd(f)$ and $\lsz_{pta}^\lfd(f)$, 
   there is a more economic way to convey the physics information on global behavior.
   Indeed, the above characteristics provide answer to the following relevant 
   question. Given arbitrary degree of saturation $\cml^\lfd$ ($0 \le \cml^\lfd \le 1$), 
   what are the values of $\lsz_{max}^\lfd$, $\lsz_{pta}^\lfd$ characterizing 
   the global behavior of the corresponding fraction support where fields are 
   concentrated? In case of a Schr\"odinger wave function we could ask what are 
   the global characteristics of the minimal region of space comprising 
   e.g. $90$\% chance to find the particle. We can express this information directly 
   by eliminating $f$ in favor of $\cml^\lfd$. In generic case this can be done 
   since $f \mapsto \cml^\lfd(f)$ is a one-to-one map of the interval $[0,1]$ 
   to itself. Formally, we are introducing the change of variable      
   \begin{equation}
      \lsz_{max}^\lfd(f) \;\longrightarrow\; \lsz_{max}^\lfd(f(\cml^\lfd)) 
      \qquad\quad
      \lsz_{pta}^\lfd(f) \;\longrightarrow\; \lsz_{pta}^\lfd(f(\cml^\lfd))
      \label{eq:50}
   \end{equation}
   While the functional forms are obviously different after the change of variable, 
   we will keep the same notation, namely $\lsz_{max}^\lfd(\cml)$ and 
   $\lsz_{pta}^\lfd(\cml)$, where it is implicitly assumed that $\cml$ is related
   to local field $\lfd(x)$. We also note that for many relevant questions $\cml$ 
   is indeed a more physical variable in the sense that $\cml$ (rather than $f$) 
   should be fixed when approaching a continuum limit. 

   It is straightforward to see how the typical localized, delocalized and global 
   physical behavior reflects in the functions $\lsz_{max}^\lfd(\cml)$ and 
   $\lsz_{pta}^\lfd(\cml)$. In Fig.~3 we plot these functions corresponding
   to cases shown in Fig.~2.

    \begin{figure}
    \begin{center}
      \centerline{
      \epsfxsize=16.5truecm\epsffile{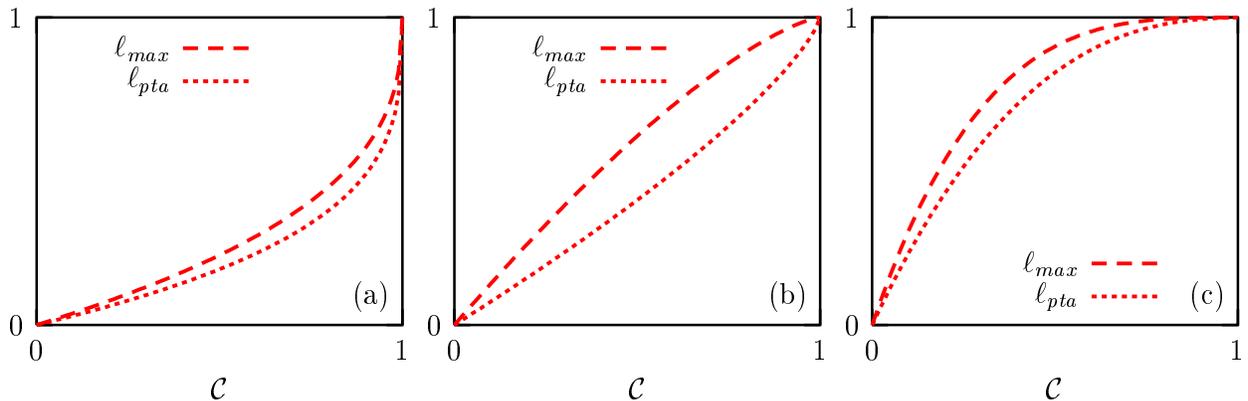}
      }
      \caption{The signatures of localized (a), delocalized (b), and inherently global (c)
               behavior.}
      \label{cartoon3}
      \vskip -0.3in
    \end{center}
    \end{figure}

\subsection{Partitioning of the fraction support}
\label{agrad}

   In our discussion so far, the space-time structure was introduced solely via the
   behavior of $\lin^\lfd(f)$, and the resulting geometric properties of fraction
   supports. However, there could be additional physics aspects of local observable 
   $\lfd(x)$, which should be taken into account when judging the global properties.
   A typical example is represented by topological charge density $q(x)$, whose
   sign has important physical consequences. Indeed, the quarks of preferential 
   chirality are attracted to regions of definite {\em sign}. This is expected 
   to be very relevant for understanding various aspects of hadron structure and 
   hadron propagation. In such a case it is then appropriate to partition the fraction 
   support accordingly, and to consider global properties of the physically relevant 
   partitioned sets~\cite{Hor02B,Hor03A,Hor02A}. In general case, we will thus consider 
   the sets $\cS^\lfd_r(f)$, $r=1,\ldots,R$, such that $\cup_r \cS^\lfd_r(f)=\cS^\lfd(f)$,
   and $\cS^\lfd_r(f) \cap \cS^\lfd_s(f) = \emptyset$ for all $r$, $s$. There are 
   several things that should be pointed out with regard to this situation.
   \smallskip

   \noindent (1) The cumulative function $\cml^\lfd(f)$ is not affected by partitioning
                 and is defined in the usual way relative to $\cS^\lfd(f)$.
   \smallskip 

   \noindent (2) The global characteristics $\lsz_{max}^{\lfd,r}(f)$, 
                 $\lsz_{pta}^{\lfd,r}(f)$ will characterize the partitioned sets and 
                 the only generic property that will change in this regard is that 
                 they do not necessarily approach unity as $f \rightarrow 1$. In 
                 principle then
                 the values $\lsz_{max}^{\lfd,r}(1)$ and $\lsz_{pta}^{\lfd,r}(1)$ are 
                 relevant characteristics that enter the classification of global 
                 properties. It is straightforward to extend our discussion to include 
                 them. 
                 \footnote{We note that in cases relevant to QCD vacuum structure 
                 that we have studied so far, we have always observed that 
                 $1=\lsz_{max}^r(1) \approx \lsz_{pta}^r(1)$, 
                 and so it is not necessary to do this explicitly here.}
   \smallskip

   \noindent (3) If there is a symmetry in the theory which relates the average 
                 properties of partitioned fraction supports (which is usually
                 the case), then the global characteristics for different $r$ 
                 just represent samples (possibly correlated) for the unique
                 functions $\lsz_{max}^\lfd(f)$ and $\lsz_{pta}^\lfd(f)$.

\subsection{Loose ends}
\label{loose}

   Our discussion in the main part of this section concentrated on clear exposition
   of basic ideas in our approach, but left open certain issues that arise in special 
   cases. In what follows we will consider some of them.

   The first is the definition of the fraction support (at the regularized level) 
   if some functional values of $\lin^\lfd(x)$ are degenerate. 
   \footnote{This will not be encountered in practice at all, but it is still 
   appropriate to complete the framework by discussing this situation.}
   The extreme case is that of a constant field $\lfd(x)$. The problem is how to 
   assign the ranking within the degenerate subset, so that the associated fraction 
   supports lead to unique definition of $\cml^\lfd(f)$, $\lsz_{pta}^\lfd(f)$ 
   and $\lsz_{max}^\lfd(f)$. Note that this is not a problem for definition 
   of $\cml^\lfd(f)$ and arbitrary ad hoc rule leads to the same function. However, 
   the functions $\lsz_{pta}^\lfd(f)$ and $\lsz_{max}^\lfd(f)$ can in principle depend 
   on the rule chosen.
   \footnote{In fact, the continuum limit of these functions will not depend on the 
   rule for ranking within degenerate subsets if these subsets become of measure zero 
   in the continuum limit.} 
   It is straightforward to complete the definition of these functions in arbitrary 
   degenerate case, so that the notion of localization, delocalization and global 
   behavior (as defined here) will correspond to its intended meaning. To do that, 
   note that the process of ranking the space-time points corresponds to specifying 
   a non-repeating sequence $x_1,x_2,\ldots,x_{N(\Omega)}$. Denoting the sets of 
   points contained in partial
   sequences containing $n$ points as $\cP^n=\{x_1,x_2,\ldots,x_n\}$, assume that
   this partial sequence has been assembled. Then the rules for continuing the
   sequence recursively are the following: (1) If the next value of intensity, i.e. 
   $\lin^{\max}_n \equiv \max \{\,\lin^\lfd(x), x \in \Omega - \cP^n \,\}$ is 
   non-degenerate, then $x_{n+1}=y$ such that $\lin^\lfd(y)=\mu^{\max}_n$.
   (2) If $\mu^{\max}_n$ is $k$-fold degenerate, i.e. if there are points
   $y_1,y_2,\ldots,y_k$ such that 
   $\lin^{\max}_n=\lin^\lfd(y_1)=\ldots = \lin^\lfd(y_k)$, then we need to specify
   the permutation $p(i), i=1,\ldots,k$ such that $x_{n+i}=y_{p(i)}$. We require 
   this permutation to be chosen in such a way that $\lsz_{max}^\lfd(f)$ grows 
   fastest for arguments $f_m\equiv m/N(\Omega), \, m\in \{n+1,n+2,\ldots,n+k\}$.
   More precisely, we require that 
   (i) $\lsz_{max}^\lfd(f_{n+1})-\lsz_{max}^\lfd(f_{n})$ is maximal over all 
       permutations. 
   (ii) If this does not specify permutation $p$ uniquely, we consider the restricted 
        degenerate set of permutations and further require that 
        $\lsz_{max}^\lfd(f_{n+2})-\lsz_{max}^\lfd(f_{n+1})$ is maximal. 
   (iii) This recursion continues and stops when $p$ is uniquely determined. 
   (iv) If $p$ is not uniquely determined at the end of the finite recursion, we 
        choose it from the remaining degenerate set by arbitrary fixed rule. One can 
        easily check that the resulting $\lsz_{max}^\lfd(f)$ and $\lsz_{pta}^\lfd(f)$ 
        are invariant under this remaining freedom. Note also that the above rules 
        lead to classification of the constant field $\lfd(x)$ as global configuration 
        which is appropriate.

   The second class of situations which needs to be mentioned is exemplified by 
   the following. Assume the existence of a strict support which becomes the set 
   of measure zero in the continuum limit. For example, a strict support could form
   form a finite low-dimensional manifold. In that case $\cml^\lfd(f)\equiv 1$ in 
   the continuum limit, and thus $\cml^\lfd(f)$ will not carry any information about 
   the behavior within the support. In that case one has to define the cumulative function 
   and the corresponding global characteristics for the support itself. In the 
   other extreme case the global characteristics $\lsz_{max}^\lfd(f)$ or 
   $\lsz_{pta}^\lfd(f)$ could become identically one in the continuum limit, 
   indicating a global behavior on sets of lower dimension. These situations are 
   of particular interest for QCD vacuum case~\cite{Hor03A}, and will be addressed 
   in a separate publication. 

\section{Cases of physical interest}
\label{cases}

   In this section we will discuss in more detail few basic examples of local observables
   $\lfd(x)$ which are of interest for studies of QCD vacuum structure, and which 
   will be analyzed in the upcoming publications. In particular, we will mention
   observables related to gauge fields $U_\mu(x)$ and the Dirac eigenmodes 
   $\psi_\lambda(x)$. To ease the notation, from now on we will skip labels of $\lfd(x)$ 
   on various characteristics discussed. The appropriate association will be obvious 
   from the context. In this section we will also give examples (in case of Dirac 
   eigenmodes) of how the methods developed here work in QCD.

\subsection{Gauge fields}
\label{gauge}    

   In pure glue lattice gauge theory all local observables are constructed from 
   fundamental gauge fields $U_\mu(x)\equiv e^{i g A_\mu(x)}$. The intensity
   of gauge fields is physically associated with the corresponding field-strength 
   tensor $U_\mu(x) \rightarrow F_{\mu\nu}(x)$ describing chromo-electric and 
   chromo-magnetic fields. Two basic physically relevant composite fields 
   characterizing the intensity of $F_{\mu\nu}$ are the scalar density 
   $s(x)\propto {\tt tr} F_{\mu\nu}F_{\mu\nu}(x)$ (action density) and 
   the pseudoscalar density $q(x)\propto {\tt tr} F_{\mu\nu}\tilde F_{\mu\nu}(x)$ 
   (topological density).
   Understanding the space-time structure of $s(x)$ and $q(x)$ in configurations
   dominating the QCD path integral is one of the most basic ingredients for
   understanding the QCD vacuum structure. 

   To characterize the global properties of this structure according to the scheme
   developed in previous sections, we first have to specify the corresponding 
   physically relevant intensities. Since $s(x)$ is positive semidefinite, it can 
   serve as its own intensity $\lin(x)$. One might be tempted to consider other 
   positive-semidefinite functions of $s(x)$, e.g. the powers 
   $\mu(x)=s^\alpha(x)$ for $\alpha>0$. The cumulative functions $\cml(f)$ would 
   obviously be different for different values of $\alpha$. However, while 
   the intensity corresponding to $\alpha=1$ has a well-defined physical meaning 
   (we study the space-time distribution of the action), this is not so for other 
   values of $\alpha$. Similarly, for $q(x)$ the appropriate intensity to use is 
   $\lin(x) \equiv |q(x)|$, since it allows for relevant comparison to the behavior 
   of $s(x)$.
 
   In the case of $q(x)$ there is a more physically-motivated approach suggested by 
   the fact that global fluctuations of topological charge are directly related to the
   $\eta'$ mass via Witten--Veneziano relation~\cite{WitVen}. One would thus like 
   to judge the global properties of the relevant space-time structure relative to 
   the saturation of topological susceptibility. While conceptually this conforms to
   the general strategy developed here, it requires some technical modifications which 
   we now describe.

   The main change one has to deal with is the definition of the cumulative function
   $\cml(f)$ because there is no intensity $\lin(x)$, local in the gauge fields, 
   that one could use to define this function directly using Eq.~(\ref{eq:30}). 
   Nevertheless, assuming that a well-defined topological density operator
   (such as the operator associated with Ginsparg-Wilson fermions~\cite{Has98A, NarNeu95})
   is at our disposal, it is still possible to construct the appropriate cumulative 
   function which will play the corresponding role in classifying the global behavior
   of the space-time structure. Indeed, defining the fraction supports $\cS(f)$ relative 
   to $\lin(x) = |q(x)|$, let us consider the corresponding cumulative charge $Q(f)$, 
   and the associated fluctuation $\chi(f)$ which is only defined in the ensemble average
   \begin{equation}
      \chi(f) \,\equiv\, \frac{\langle\, Q^2(f) \,\rangle \,-\, 
                               \langle\, Q(f)   \,\rangle^2}{V}
      \qquad\qquad
      Q(f) \,\equiv\, \sum_{x\in \cS(f)} q(x)     \qquad\quad
      \label{eq:55}
   \end{equation}
   where $V=a^4 N(\Omega)$ is the physical volume. We then define
   the normalized cumulative function of global fluctuations via
   \begin{equation}
      \cml(f) \;\equiv\; \chi(f)/\chi(1)
   \end{equation}

   There are two points that need to be emphasized here. (1) While we call the $\cml(f)$
   defined above a cumulative function in analogy with definition in Eq.~(\ref{eq:30}), it
   is not a priori obvious that it has the specific properties of the cumulative function.
   In particular, it is not obvious that it is a concave non-decreasing function. Indeed,
   we are making an assumption that the ordering induced by $|q(x)|$ will ensure these 
   properties in the ensemble average. This is reasonable since it is expected that regions 
   with most intense fields will contribute most to global fluctuations. Our numerical data 
   a posteriori confirm that this is indeed the case in QCD~\cite{Hor04B}. (2) The precise 
   determination of $\cml(f)$ does not necessarily require the use of very large ensembles 
   to compute the physical susceptibility $\chi(1)$ to comparable accuracy. Indeed, computing 
   $\cml(f)$ directly via jackknife samples of $\chi(f)/\chi(1)$ leads to much smaller 
   fluctuations over the whole range of $f$~\cite{Hor04B}.

\subsection{QCD Dirac Zero Modes}
\label{zmodes}    

   For Dirac eigenmodes $\psi_\lambda(x)$ 
   ( i.e. $D \psi_\lambda (x) = \lambda \psi_\lambda(x)$ )
   the basic scalar and pseudoscalar composites
   are the ``density'' $d_\lambda(x) \equiv \psi_\lambda^\dagger(x) \psi_\lambda(x)$
   and ``chirality'' $c_\lambda(x) \equiv \psi_\lambda^\dagger(x) \gfive \psi_\lambda(x)$.
   However, the zero modes are globally chiral and this is exactly true even on the lattice 
   if overlap fermions are used (which we assume). We thus only have one independent 
   bilinear to study, which we choose to be the density $\lfd(x) \equiv d(x)$.
   In direct analogy to the case of Schr\"odinger particle, the associated intensity
   $\lin(x)$ for determining the fraction supports and for computing the cumulative functions 
   will be the density $d(x)$ itself. Note that the choice of the bilinear form (rather than 
   higher powers) is further physically motivated by the fact that the fermionic action is 
   bilinear in fermion fields.

   To illustrate how the global characteristics of the space-time structure introduced
   here behave in QCD, we have analyzed the low-lying eigenmodes of the overlap operator.
   In particular, we have considered an ensemble of $20$ Wilson gauge configurations 
   on $16^4$ lattice at $\beta=6.07$. This corresponds to the lattice spacing of $a=0.082$ 
   fm as determined from the string tension~\cite{Sommer}, and the linear extent 
   $L\approx 1.3$ fm. The eigenmodes of the overlap operator with negative mass $\rho=1.368$ 
   ($\kappa=0.19$) were computed using the Zolotarev approximation to implement the operator 
   numerically.  The details of this numerical implementation are given elsewhere 
   (see e.g. \cite{Hor02B,Chen03}). 

   \begin{figure}
   \begin{center}
    \vskip -0.3in
     \centerline{
     \epsfxsize=9.2truecm\epsffile{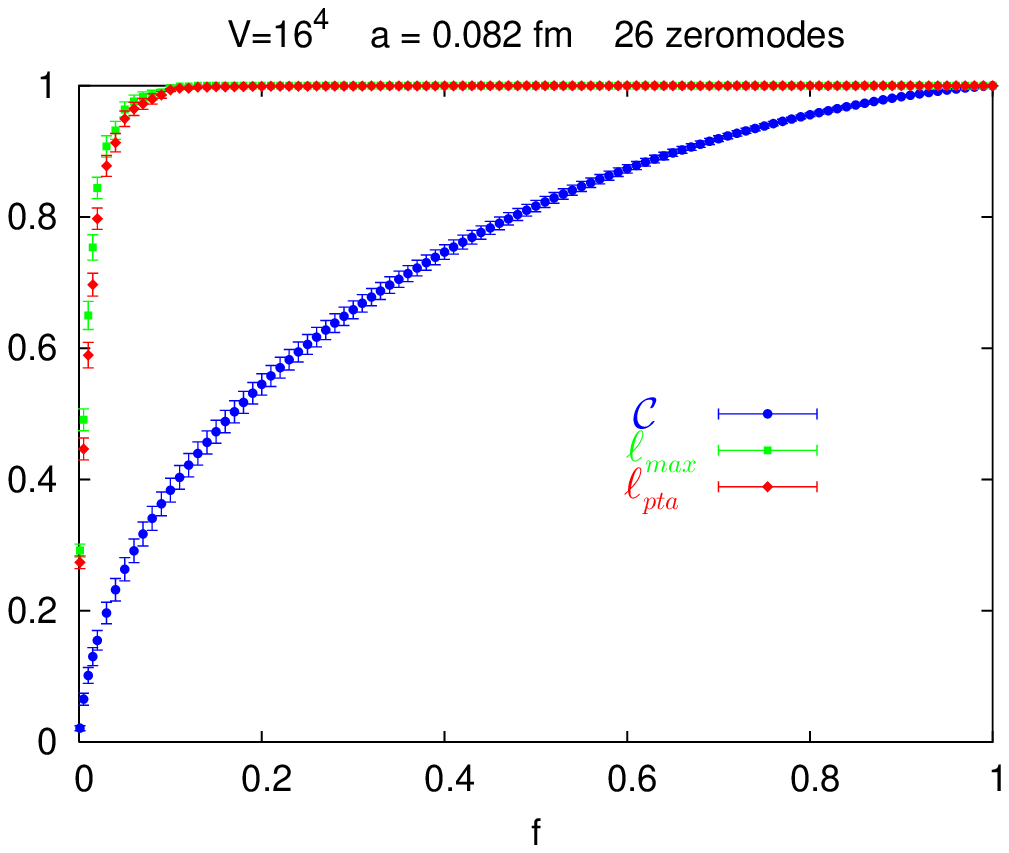}
     \hskip -0.30in
     \epsfxsize=9.2truecm\epsffile{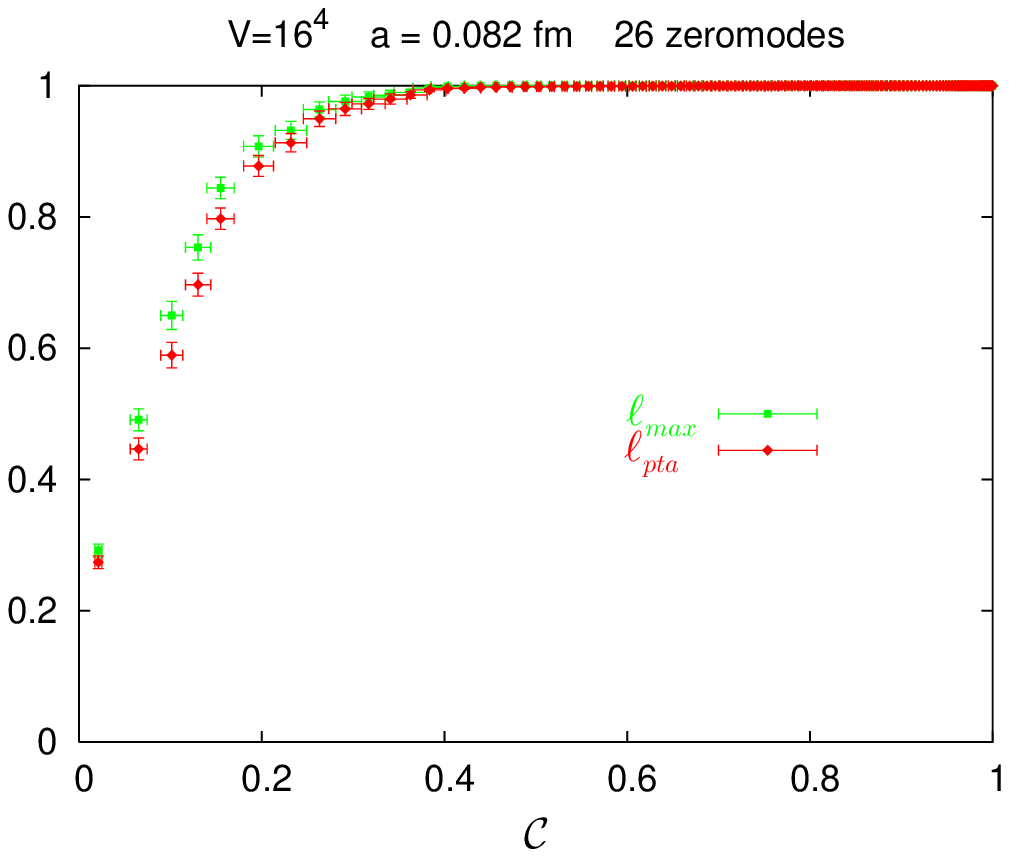}
     }
     \centerline{
     \epsfxsize=9.2truecm\epsffile{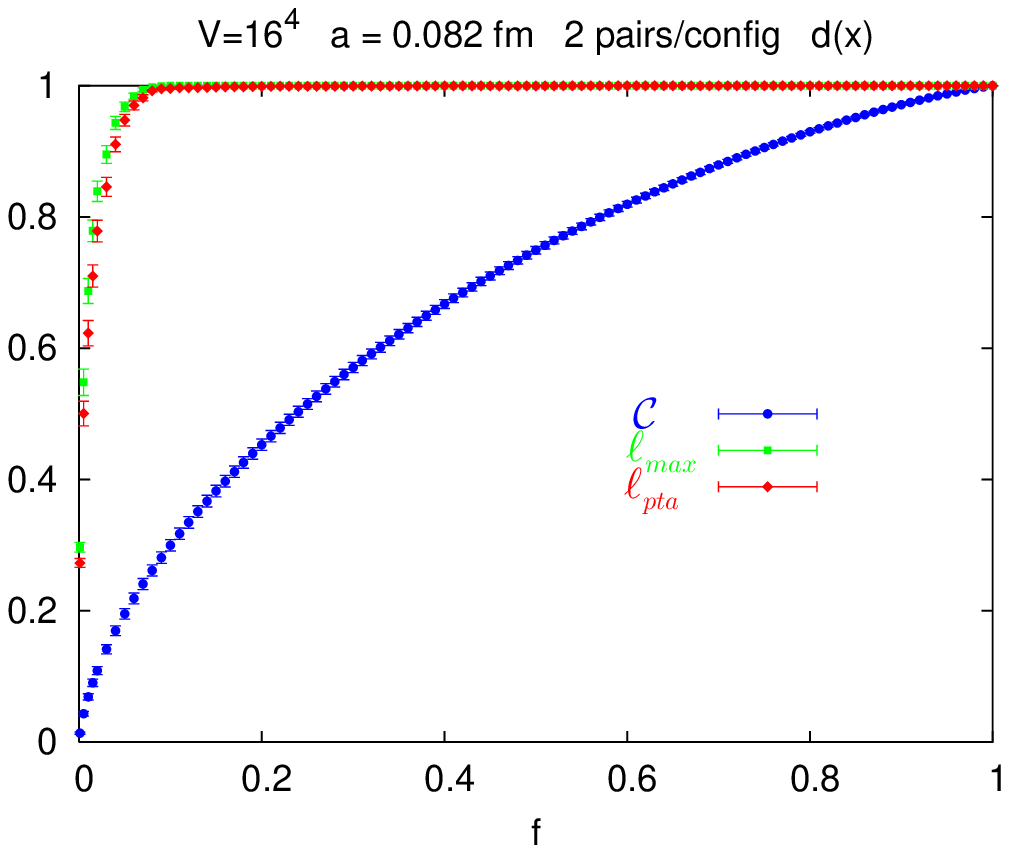}
     \hskip -0.30in
     \epsfxsize=9.2truecm\epsffile{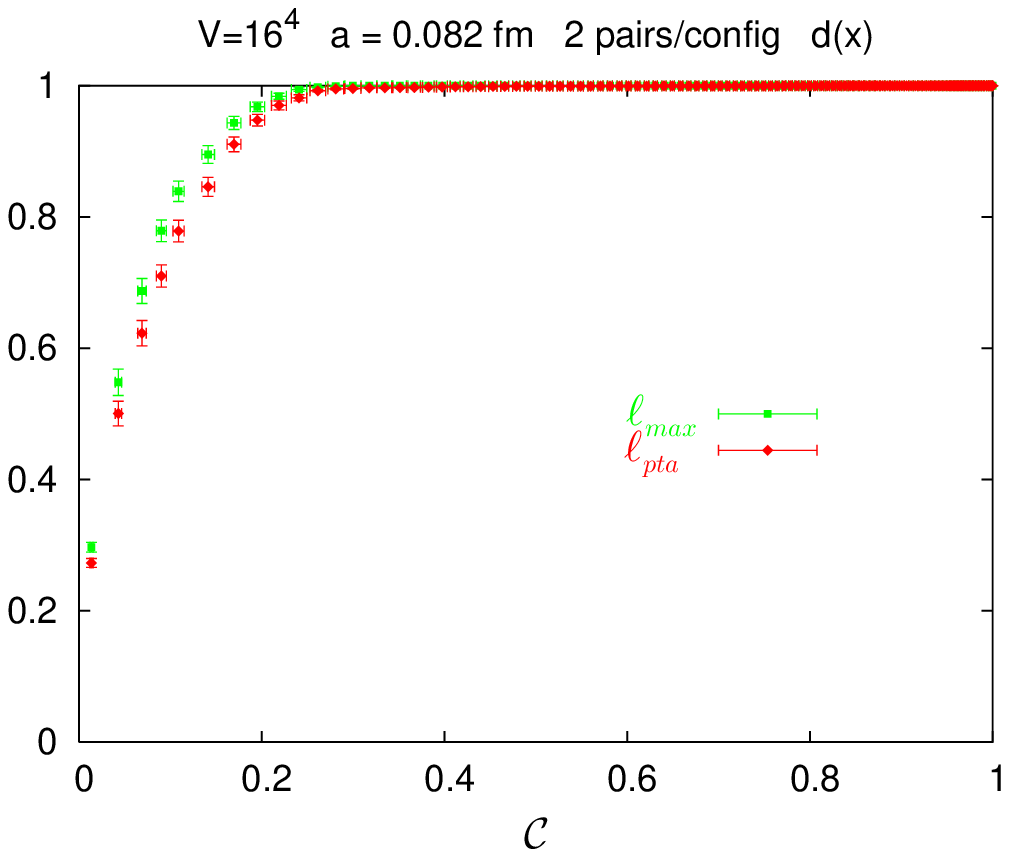}
     }
     \centerline{
     \epsfxsize=9.2truecm\epsffile{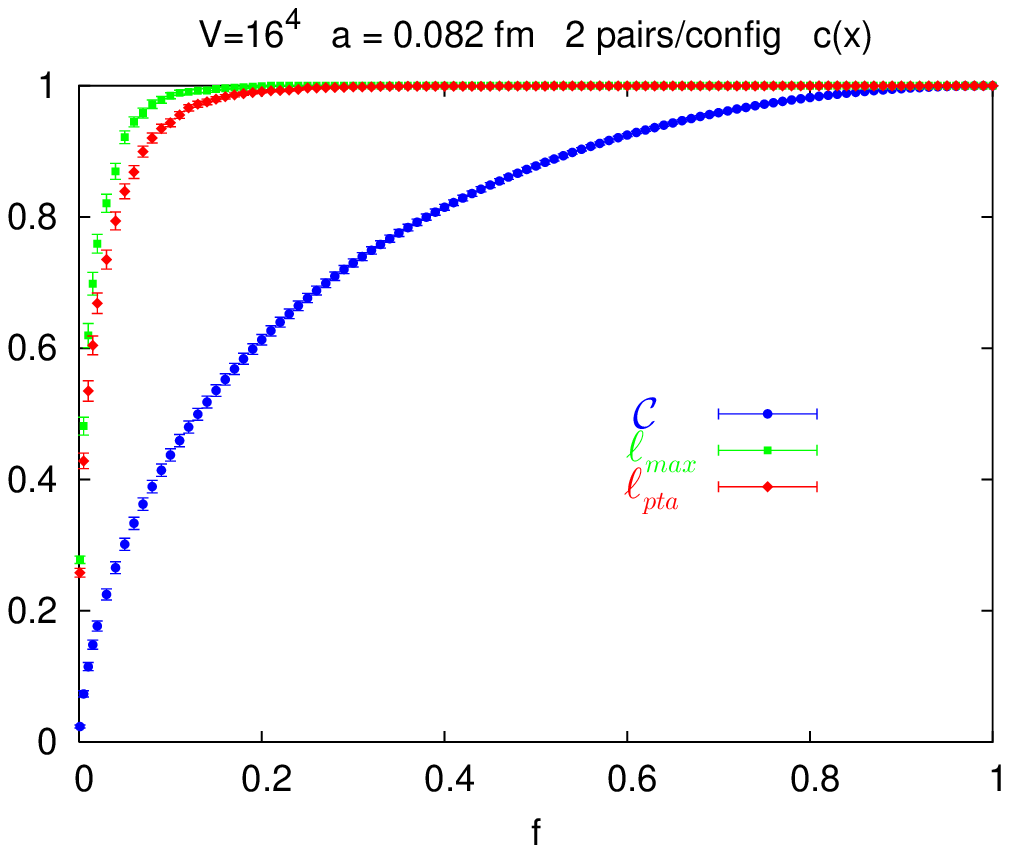}
     \hskip -0.30in
     \epsfxsize=9.2truecm\epsffile{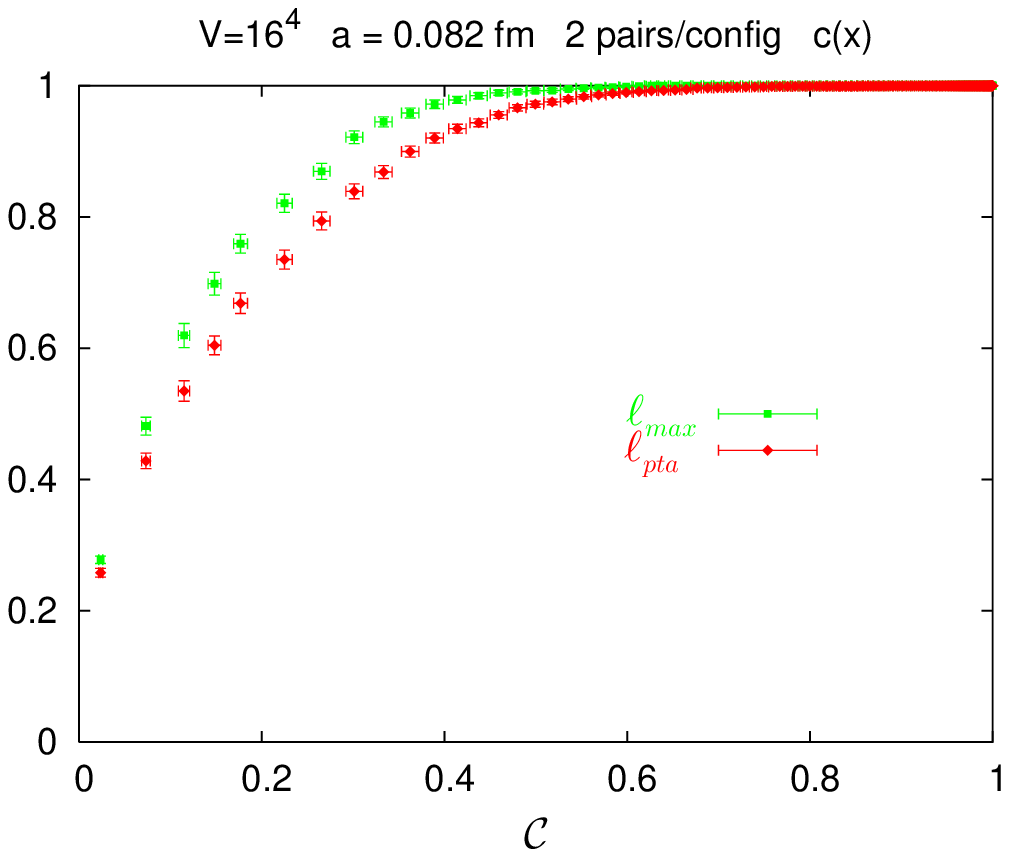}
     }
     \caption{Characteristics of global behavior for Dirac zero modes (top), Dirac near-zero modes
              via density (middle) and Dirac near-zero modes via chirality (bottom). The structure 
              is clearly super-long-distance in all cases. See the discussion in the text.}
     \label{qcdmod_fig}
   \end{center}
   \end{figure}

  We found the total of $26$ zero modes in the above ensemble and computed the average 
  $\cml(f)$, $\lsz_{max}(f)$ and $\lsz_{pta}(f)$. The results are shown in 
  Fig.~\ref{qcdmod_fig} (top left) together with the associated $\lsz_{max}(\cml)$ and 
  $\lsz_{pta}(\cml)$ (top right). One can clearly observe that the structure is geometrically 
  global at the fractions of just few percent (e.g. $\lsz_{pta}(f=0.036)=0.9$ ). At such 
  fractions the cumulative function only achieves values $\cml\approx 0.2$. Following our 
  discussion in Sec.~\ref{cumul} we thus conclude that we clearly observe a 
  {\em super-long-distance space-time structure in QCD Dirac zero modes}. Note also that 
  $\lsz_{max}(f)$ and $\lsz_{pta}(f)$ are quite close to one another, indicating that there 
  is very little fragmentation at arbitrary fraction. In fact, the structure typically comes 
  in a single dominating piece as it is grown from very small fractions up.

\subsection{QCD Dirac Near-Zero Modes}
\label{nzmodes}    

  In the case of near-zero modes we can study the global behavior in both $d(x)$ and
  $c(x)$. Due to $\gfive$-Hermiticity, the near-zero modes of the overlap operator 
  come in complex-conjugate pairs with identical $d(x)$ and $c(x)$. We will thus consider
  only a single representative of the pair. For our calculation, we have considered
  two lowest-lying pairs for every configuration from the ensemble. This gave us 
  altogether $40$ modes to analyze.

  We start with the density for which the appropriate intensity is $\lin(x) \equiv d(x)$ as in 
  the case of zero modes. The resulting global characteristics are shown in Fig.~\ref{qcdmod_fig} 
  (middle). From this data it is very clear again that {\em the scalar structure in QCD 
  near-zero modes is super-long-distance}. For example, $\lsz_{pta}(f=0.038)=0.9$, while 
  $\cml\approx 0.16$  at the same fraction. Similarly to zero modes, the functions
  $\lsz_{max}(f)$ and $\lsz_{pta}(f)$ are quite close one to another and the structure is 
  dominated by a single connected piece. As a side remark, note also that the cumulative 
  function $\cml(f)$ has steeper behavior in case of zero modes than in case of near-zero 
  modes. This results in steeper behavior of functions $\lsz_{max}(\cml)$, $\lsz_{pta}(\cml)$ 
  for near-zero modes. We will study such details in a dedicated publication~\cite{gl_modes}.

  In case of chirality the appropriate intensity to study is $\lin(x)=|c(x)|$. Since the sign
  of local chirality carries a relevant physical information (e.g. on the preferred duality 
  of the underlying gauge field and the qualitative features of the quark propagation), we
  partition the fraction supports into subsets of definite sign (see Sec.~\ref{agrad}).
  Since QCD dynamics has no preference for a particular chirality, we average over connected 
  subsets of both signs in calculation of $\lsz_{pta}(f)$. Also, in calculation of 
  $\lsz_{max}(f)$ we consider the maximum over connected subsets of both signs. The results
  are shown in Fig.~\ref{qcdmod_fig}. Again, we find the space-time characteristics 
  corresponding to clear {\em super-long distance structure in $c(x)$}. Note that the functions
  $\lsz_{pta}(f)$, $\lsz_{max}(f)$ saturate slightly slower than in the case of $d(x)$
  which is due to the partition and the fact that the connected subsets of opposite chirality
  ``obstruct'' one another in their global behavior. At the same time the cumulative function
  is steeper than in the case of $d(x)$ (it is in fact similar to cumulative function 
  for zero modes) resulting in the slower saturation of $\lsz_{max}(\cml)$ and 
  $\lsz_{pta}(\cml)$.

\section{Conclusions}

  In this paper we have proposed the framework for studying the issues of localization
  in QCD vacuum in a model-independent manner~\footnote{The methods introduced here are by no means 
  special to QCD or to four dimensions.}. This framework was set up in such a way 
  as to inquire into three hypothetical scenarios underlying the QCD dynamics. The first scenario 
  (localization) involves the existence of fundamental localized objects in the gauge
  field populating the QCD vacuum and driving the dynamics. The second scenario (delocalization)
  involves localized entities which however clump into global structures (via interaction) and 
  can support physical correlations over very large distances. A distinctive property of 
  delocalization is that the localized objects still preserve their identity and can serve as a basis 
  for physical analysis of vacuum properties. The third scenario (inherently global or 
  super-long-distance behavior) is qualitatively different in that there are fundamental entities
  that themselves exhibit global behavior. The geometric structure of these entities can facilitate 
  correlations over arbitrarily large distances and the vacuum properties can be fully understood
  only in terms of these global objects. 
   
  The approach proposed in this work is based on the idea that the existence of gauge field objects 
  with above properties will lead to a corresponding geometric behavior in local composite fields, and 
  possibly also in Dirac eigenmodes. In particular, the space-time regions containing dominating 
  fields (the ``support'' of the field) will behave in the specific geometric manner, forming localized, 
  partially localized or global geometric structure. However, since the definition of field support 
  on physical grounds is not generically available, we proposed the use of cumulative functions 
  with simultaneous monitoring of both the degree of saturation of the field, and of the global 
  characteristics of the underlying fraction supports. It was then proposed that the specific 
  functional behavior of these dependencies can be associated with the three physical situations
  described above. 

  We have discussed in detail the application of these ideas to special cases of action density, 
  topological density, and to the scalar and pseudoscalar density of Dirac low-lying modes. Our 
  approach was illustrated via the study of overlap zero modes and near-zero modes. The results clearly 
  demonstrate the super-long distance nature of the underlying gauge structure as reflected in 
  the Dirac eigenmodes. They also provide the quantitative validation for the original suggestion 
  by the Kentucky group that the Dirac near-zero modes form global structure (``ridges'' rather than 
  separate peaks)~\cite{Hor02A}.
  \footnote{We emphasize that in the case of Dirac modes we talk about the eigenmodes of the proper 
            {\em chirally symmetric} lattice Dirac operator such as the overlap operator. 
            Exponentially localized modes could be observed e.g. in hermitian Wilson-Dirac
            operator for unphysical values of the hopping parameter~\cite{Gol03}.}
            
  Finally, as emphasized in the introduction, the aim of this paper (first in the series) is to 
  introduce techniques suitable for analysis of QCD space-time structure obtained 
  in a model-independent manner. In this approach, gauge fields are not processed or altered 
  in any way, and the structure is uncovered as it appears directly in the configurations dominating 
  the QCD path integral. The culmination of these attempts so far is the discovery of the strictly 
  low-dimensional, inherently global sheet/skeleton structure in the overlap topological charge 
  density~\cite{Hor03A}.
  \footnote{It is interesting that the notion of strictly low-dimensional 
  structure is also emerging in an indirect way, using the monopole and center-vortex detection
  (projection) techniques~\cite{Zakh}. If there is a relation between these findings it is yet to be 
  found and understood.}
  The methods developed in this work (see Sec.~\ref{gauge}) provide a formal framework for a detailed
  quantitative confirmation of the super-long-distance nature of this structure~\cite{Hor04B}.

\bigskip\medskip
\noindent {\Large \bf Acknowledgments} 
\medskip

\noindent
Numerous discussions with members of the Kentucky group on the topics of this work are gratefully 
acknowledged. Thanks to Andrei Alexandru and Hank Thacker for feedback on the manuscript. The author
is indebted to Andrei Alexandru and Nilmani Mathur for help with the figures, and to Jianbo Zhang for 
the help with the data, part of which was used to illustrate the methods proposed in this manuscript. 
The Kentucky group is acknowledged for making this data available. The author benefited from 
communications with Ganpathy Murthy, {\v S}tefan Olejn\'{\i}k and Joe Straley.

\begin{appendix}

\section{Generic Definitions of the Support}
\label{app:1}

      Assume that we are given a generic function $\lfd(x)$ on space-time, and the associated 
      real-valued positive-semidefinite intensity $\lin(x)$. If the specific properties of $\lfd(x)$ 
      are not known, it is tempting to use some generic way of determining the space-time region 
      where the function is effectively concentrated, i.e. its {\em effective support}. This 
      support is implicitly associated with space-time regions occupied by strongest fields, such 
      that fields in the rest of the space-time can either be neglected or behave in a qualitatively 
      different (low intensity) manner. Thus, from the geometric point of view, effective support is 
      one of the fraction supports defined in Sec.~\ref{cumul}, and the goal of a given generic 
      procedure is to fix the corresponding fraction ({\em effective fraction $f_{eff}$}). However, 
      as emphasized in Sec.~\ref{cumul}, the generic definitions of this type are highly ambiguous. 

      To illustrate this, let us concentrate on the scalar structure of Dirac eigenmodes. The same 
      discussion can be extended straightforwardly to the case of arbitrary local observable $\lfd(x)$.
      Given a (normalized) Dirac eigenmode $\psi(x)$, the physically relevant intensity characterizing
      its (scalar) strength is $\lin(x)\equiv d(x)=\psi^{\dagger}\psi(x)$, and the definition
      of the support will thus be given in terms of this associated function $f_{eff}=f_{eff}[d(x)]$.
      Consider, for example, the following definitions of $f_{eff}$.

\begin{figure}
      \begin{center}
         \vskip -0.3in
         \centerline{
         \epsfxsize=8.8truecm\epsffile{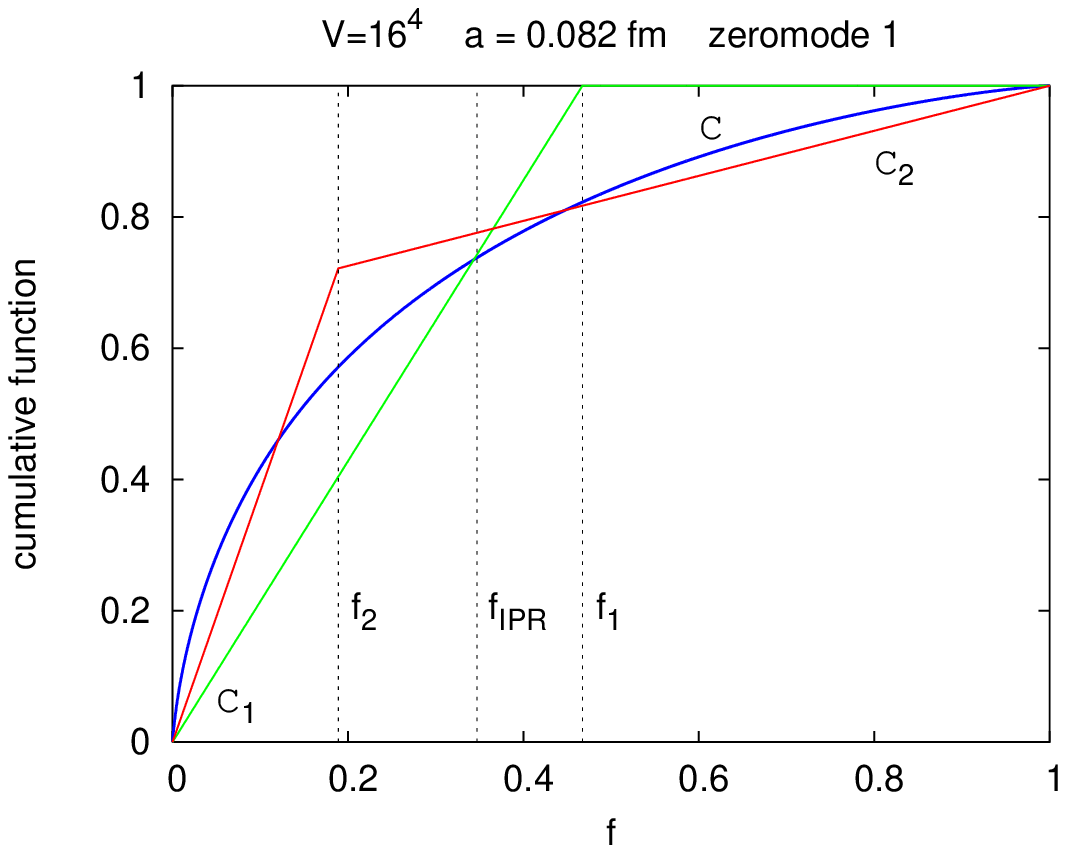}
         \hskip -0.15in
         \epsfxsize=8.8truecm\epsffile{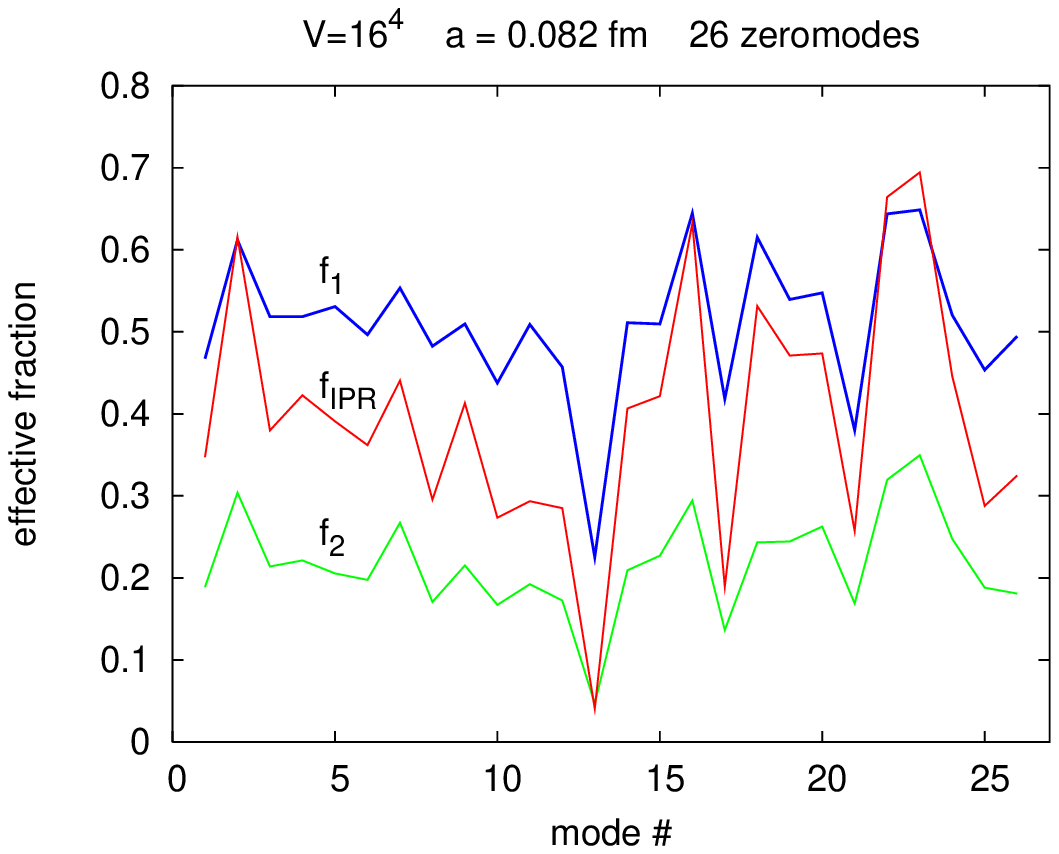}
         }
         \caption{Left: The cumulative function $\cml(f)$ for zero mode \#1 and the associated
                  effective fractions $f_1$ (with $\cml_1(f)$), $f_2$ (with $\cml_2(f)$), and
                  $f_{IPR}$ are shown. Right: Three generic definitions of effective fraction 
                  discussed in the text for all 26 zero modes.}
         \label{effrac_fig}
         \vskip -0.4in
      \end{center}
      \end{figure}

      (1) A natural possibility is to base the definition on the behavior of the corresponding 
          cumulative function $\cml(f)$. In Fig.~\ref{effrac_fig} (left) we show $\cml(f)$ for the 
          zero mode \#1 from the set analyzed in Sec.~\ref{zmodes}. Even though we work on the latticized 
          torus with $N$ sites, we will keep the ``continuous'' notation  and treat $\cml(f)$ as 
          a piece-wise linear continuous function connecting the points of the discrete map 
          (see Sec.~\ref{cumul}).
          $\cml(f)$ is a concave function on the interval $[0,1]$ such that $\cml(0)=0$ and 
          $\cml(1)=1$, and hence $1\ge I \equiv \int_0^1 df \,\cml(f) \ge 1/2$. Clearly, if $d(x)$
          is sharply concentrated on small fraction of space-time, then $I \approx 1$, and when
          $d(x)$ is spread out over the whole lattice then $I \approx 1/2$. In an extreme case
          of localization on single site we have $I=1-1/2N$ ($I=1$ in the continuum limit), and $I=1/2$ 
          for constant $d(x)$. We can thus in principle base the definition of $f_{eff}$ on the value 
          of $I$. The simplest possibility is to use the linear expression in $I$ which interpolates 
          between the above extreme values, i.e.
          \begin{equation}
             f_{eff} \,\equiv 2 ( 1-I ) \,\equiv\, f_{1}
             \label{eq:60}
          \end{equation}
          This definition gives $f_1=1/N$ and $f_1=1$ for $d(x) = \delta_{x,y}$ and 
          $d(x) = 1/N$ respectively. Also, $0 \le f_1 \le 1$ for arbitrary $d(x)$, 
          and thus $f_1$ has the intended meaning of the effective fraction. It is ``tailored'' for 
          the functions $d(x)$ that are constant on fraction $f_1$ of sites and vanish on the rest 
          of the sites. In effect then, the above procedure associates with arbitrary $d(x)$ a unique 
          function of this type, such that the corresponding integrals $I$ are the same. 
          In Fig.~\ref{effrac_fig} (left) we show the value $f_1$ together with the cumulative function 
          $\cml_1$ for the associated constant function on the strict support. 

      (2) Other than its simplicity, there is no particular reason to choose $f_1$ to be the definition
          of effective fraction in a generic situation. For example, instead of relating $d(x)$ to the 
          constant function on strict support, we could relate it to the function which is constant 
          and non-zero both on the support and (with lower intensity) on the complement of the support. 
          To uniquely specify such function (and with it the corresponding fraction $f_{eff}\equiv f_2$), 
          we can e.g. require simultaneously that (a) the integrals $I$ of the cumulative functions are 
          the same and that (b) the corresponding $\cml_2(f)$ is the best approximation of $\cml(f)$, 
          i.e. that $\int_0^1 df \, (\cml(f) - \cml_2(f))^2$ is minimal. With this choice the resulting 
          $f_2$ and $\cml_2(f)$ for zero mode \# 1 are shown in Fig~\ref{effrac_fig} (left). Note that 
          this definition could be useful if it is {\em a priori known} that the actual space-time 
          distributions in question have a well-defined, approximately constant core, but also an 
          approximately constant background. Again, in the extreme cases of $d(x) = \delta_{x,y}$ and 
          $d(x) = 1/N$, we obtain $f_2=1/N$ and $f_2=1$ respectively, while $0 \le f_2 \le 1$ for
          arbitrary $d(x)$.

      (3) Effective fraction can also be based on the inverse participation ratio (IPR)~\cite{Thou74}. 
          This technique was developed in condensed matter physics where both its virtues and limitations 
          are well known. Its original intended use was to monitor the departures from exponential 
          localization. 
          \footnote{In lattice QCD literature the IPR-s started to be used in the study of 
          Dirac eigenmodes only relatively recently~\cite{Gat01A}. (See also a new conference 
          proceedings preprint~\cite{Aub04}, where they are employed in an attempt to estimate
          the dimensionality of structures in the eigenmodes of the Asqtad operator.)} 
          In this case the effective fraction can be assigned generically via 
          \begin{equation}
             \frac{1}{f_{eff}} \,\equiv \, \frac{1}{f_{IPR}} \,\equiv\, N \sum_x\, d^2(x)
          \end{equation}      
          Needless to say, in this case we have also that $f_{IPR}=1/N$ and $f_{IPR}=1$
          for $d(x) = \delta_{x,y}$ and $d(x) = 1/N$ respectively. Also, $0 \le f_{IPR} \le 1$ in 
          generic case.  

       As can be seen clearly on the example of the overlap Dirac zero mode in Fig.~\ref{effrac_fig}
       (left), the above generic definitions, while quite sensible for their designed purpose, 
       give wildly different estimates for what the effective fraction is. It is thus potentially 
       dangerous to use generic methods to determine the effective fraction in situations where 
       sufficiently detailed characteristics of typical space-time distributions are not known, 
       while the quantitative aspects of the problem are important. In particular, it would not be 
       possible to seriously address the issues of global behavior in QCD vacuum at this stage 
       by using a particular definition from the infinite set of generic possibilities available. 
       At the same time, using a more complete information stored in the cumulative function
       makes it possible to resolve this issue.

       Finally, we should emphasize that generic definitions, such as the above, are rather robust 
       for comparative purposes, i.e. to decide if the support of mode $\psi_1$ is larger (more potent)
       than the support of mode $\psi_2$. Indeed, in Fig.~\ref{effrac_fig} (right) we show the effective
       fractions for all $26$ modes using the above three definitions, and the comparisons are largely
       consistent.

\end{appendix}


\begin{thebibliography}{99}


\bibitem{Hor02B} I.~Horv\'ath, S.J.~Dong, T.~Draper, F.X.~Lee, K.F.~Liu, H.B.~Thacker, J.B.~Zhang, 
                 Phys.~Rev. {\bf D67}, 011501(R) (2003);
                 I.~Horv\'ath \emph{et al}., Nucl. Phys. {\bf B}  
                 (Proc. Suppl.) 119, 688 (2003);

  
\bibitem{Hor03A} I.~Horv\'ath, S.J.~Dong, T.~Draper, F.X.~Lee, K.F.~Liu, N.~Mathur, H.B.~Thacker, 
                 J.B.~Zhang, Phys.~Rev. {\bf D68}, 114505 (2003);
                 I.~Horv\'ath \emph{et al}., Nucl. Phys. {\bf B}  
                 (Proc. Suppl.) 129\&130, 677 (2004);
                 I.~Horv\'ath \emph{et al}., Proceedings of the ``Quark Confinement and
                 the Hadron Spectrum V", Gargnano, Italy, Sep 10-14, 2002, N.~Brambilla
                 and G.~Prosperi editors, World Scientific(2003) p.312,
                 {\tt hep-lat/0212013}.

\bibitem{Hor01A} I.~Horv\'ath, N.~Isgur, J.~McCune, H.B.~Thacker,
                 Phys.~Rev.~{\bf D65}, 014502 (2002).  

\bibitem{deG01A} T.~deGrand, A.~Hasenfratz, Phys.~Rev. {\bf D64}, 034512 (2001).        

\bibitem{Hor02A} I.~Horv\'ath, S.J.~Dong, T.~Draper, N.~Isgur, F.X.~Lee, K.F.~Liu, J.~McCune, 
                 H.B.~Thacker, J.B.~Zhang, Phys.~Rev. {\bf D66}, 034501 (2002).

\bibitem{Followup} 
         T. DeGrand, A. Hasenfratz, Phys. Rev. {\bf D65}, 014503 (2002);
         I. Hip {\em et al}., Phys. Rev. {\bf D65}, 014506 (2002);
         R. Edwards, U. Heller, Phys. Rev. {\bf D65}, 014505 (2002);
         T. Blum {\em et al}., Phys. Rev. {\bf D65}, 014504 (2002);
         C. Gattringer {\em et al}., Nucl. Phys. {\bf B618} (2001) 205.

\bibitem{Other} 
         N.~Cundy, M.~Teper, U.~Wenger, Phys.~Rev. {\bf D66}, 094505 (2002);
         C.~Gattringer, Phys. Rev. Lett. {\bf 88} 221601 (2002);
         P.~Hasenfratz {\em et al}., Nucl. Phys. {\bf B643} (2002) 280;
         C.~Gattringer, Phys.~Rev. {\bf D67}, 034507 (2003);
         C.~Gattringer and R.~Pullirsch, Phys.~Rev. {\bf D69}, 094510 (2004).
         T.~Draper {\em et al}., {\tt hep-lat/0408006}.

\bibitem{overlap} H.~Neuberger, Phys.~Lett {\bf B417} (1998) 141; 
                                Phys.~Lett.{\bf B427} (1998) 353.

\bibitem{Tha04} H.~Thacker, S.~Ahmad and J.~Lenaghan, {\tt hep-lat/0409079}.

\bibitem{And58} P.~Anderson, Phys.~Rev. {\bf 109}, 1492 (1958).

\bibitem{gl_modes} I.~Horv\'ath \emph{et al}., in preparation.

\bibitem{WitVen} E.~Witten, Nucl. Phys. B{\bf 156}, 269 (1979);
                 G.~Veneziano, Nucl. Phys. B{\bf 159}, 213 (1979).
         
\bibitem{Has98A}
   P. Hasenfratz, V. Laliena, F. Niedermayer, Phys.~Lett. {\bf B427} (1998) 125.

\bibitem{NarNeu95} R.~Narayanan and H.~Neuberger, Nucl.~Phys. {\bf B443}, 305 (1995).

\bibitem{Hor04B} I.~Horv\'ath, Talk at the workshop ``The QCD Vacuum from a Lattice
                 Perspective'', Jul 29--31 2004, Regensburg;
                 I.~Horv\'ath \emph{et al}., in preparation.

\bibitem{Sommer} M.~Guagnelli, R.~Sommer, H.~Wittig, Nucl.~Phys. {\bf B535} (1998) 389.

\bibitem{Chen03} Y.~Chen \emph{et al}., Phys.~Rev. {\bf D70}, 034502 (2004).

\bibitem{Gol03} M.~Golterman and Y.~Shamir, Phys.~Rev. {\bf D68}, 074501 (2003).  

\bibitem{Zakh} F.V.~Gubarev, A.V.~Kovalenko, M.I.~Polikarpov, S.N.~Syritsyn
                and V.I.~Zakharov, Phys.~Lett. {\bf B574}, 136 (2003);
                A.V.~Kovalenko, M.I.~Polikarpov, S.N.~Syritsyn, V.I.~Zakharov, 
                {\tt hep-lat/0408014}.

\bibitem{Thou74} D.J.~Thouless, Phys.~Rep. {\bf 13} (1974) 93.

\bibitem{Gat01A} C.~Gattringer {\em et al}. Nucl.~Phys. {\bf B617}, (2001) 101.

\bibitem{Aub04} C.~Aubin {\em et al}. [MILC], {\tt hep-lat/0410024}.

\end{thebibliography}
\end{document}
\bye